\documentclass[preprints,review,accept,moreauthors,pdftex]{Definitions/mdpi}
\firstpage{1} 
\pubvolume{1}
\issuenum{1}
\articlenumber{0}
\pubyear{2021}
\copyrightyear{2020}
\datereceived{} 
\dateaccepted{} 
\datepublished{} 
\hreflink{https://doi.org/} 

\definecolor{salmon}{rgb}{0.95,0.5,0.25}
\definecolor{olive}{rgb}{0.5,0.7,0.25}

\Title{The cusp-core problem in gas-poor dwarf spheroidal galaxies}

\TitleCitation{The cusp-core problem in gas-poor dwarf spheroidal galaxies}

\Author{Pierre Boldrini $^{1},^{2}$}

\AuthorNames{Pierre Boldrini}

\AuthorCitation{Boldrini, P.}

\address{%
$^{1}$ \quad Universit́e de Lorraine, CNRS, Inria, LORIA, F-54000 Nancy, France\\
$^{2}$ \quad Sorbonne Universit\'e, CNRS, UMR 7095, Institut d'Astrophysique de Paris, 98 bis bd Arago, 75014 Paris, France}

\corres{Correspondence: boldrini@iap.fr}

\abstract{This review deals with the inconsistency of inner dark matter density profiles in dwarf galaxies, known as the cusp-core problem. Particularly, we aim to focus on gas-poor dwarf galaxies. One of the most promising solutions to this cold dark matter small scale issue is the stellar feedback but it seems to be only designed for gas-rich dwarfs. However, in the regime of classical dwarfs, this core mechanism becomes negligible. Therefore, it is required to find solutions without invoking these baryonic processes as dark matter cores tend to persist even for these dwarfs, which are rather dark matter-dominated. Here we have presented two categories of solutions. One consists of creating dark matter cores from cusps within cold dark matter by altering the dark matter potential via perturbers. The second category gathers solutions which depict the natural emergence of dark matter cores in alternative theories. Given the wide variety of solutions, it becomes necessary to identify which mechanism dominates in the central region of galaxies by finding observational signatures left by them in order to highlight the true nature of dark matter.}


\DeclareUnicodeCharacter{0301}{\'{e}}

\begin{document}

\section{Introduction}

\subsection{Lambda Cold Dark Matter paradigm and its dark matter cusps}

The nature of dark matter (DM) is currently one of the most fundamental and elusive mysteries in physics. One way to constraint the nature of the DM is to understand how DM is distributed in galaxies. The DM is arranged, more particularly in the centre of galaxies, according to the properties that we attribute to it. In the prevailing cosmological theory, Lambda Cold Dark Matter ($\Lambda$CDM), a collisionless and non-relativistic non-luminous matter spans our entire Universe \citep{1984Natur.311..517B}. However, DM could be more complex and hotter than simple CDM. Indeed, DM could be an ultralight scalar field or self-interacting or have several components. Nevertheless, the CDM paradigm can provide a quantitative description of the Universe at present and is extremely successful at explaining the Universe on large scales \citep{2003ApJS..148..175S,2002ApJ...581...20C} and also many important aspects of galaxy formation \citep{2006Natur.440.1137S,2011ApJ...742...16T}.

\medskip

CDM cosmological simulations including only DM particles predict that DM halos should have density profiles that behave as $r^{-1}$ at small radii. Halo mergers gradually drive the halo density profiles towards a central density cusp with a sharp decline towards their outskirts \citep{1988ApJ...327..507F,1991ApJ...378..496D,1994ApJ...436..467G}. These early simulations of structure formation found a universal cuspy density profile in halos ranging from dwarf galaxies to galaxy clusters \citep{1996ApJ...462..563N}. This density profile, which is almost independent of halo mass, cosmological parameters and the power spectrum of initial fluctuations, appeared to be well-described by the following form \citep[][hereafter NFW]{1996ApJ...462..563N,1997ApJ...490..493N}:
\begin{equation}
    \rho_{\rm NFW}(r)=\frac{\rho_{0}}{\left(\frac{r}{r_{\mathrm{s}}}\right)\left(1+\frac{r}{r_{\mathrm{s}}}\right)^{2}} ,
    \label{eq1}
\end{equation}
where $r$ is the distance from the centre of the DM halo, and $\rho_{0}$ and $r_{\mathrm{s}}$ represent the central density and scale radius, respectively. The NFW profile is a double power-law that transitions from $r^{-1}$ at small radii to $r^{-3}$ at large radii (see Equation~\eqref{eq1}). The scale radius marks the transition between the two slopes in the NFW profile. As the NFW profile appears to be the generic consequence of halo mergers and becomes more resilient, this might explain why this density profile is universally observed in most cosmological simulations. Nevertheless, later studies show that the DM density profile does not seem to be universal. Indeed, DM density profiles rise steeply at small radius more like $\rho(r)\propto r^{-\alpha}$ with $\alpha=$ 0.8 - 1.4 \citep{1997ApJ...477L...9F,1998ApJ...499L...5M,2010MNRAS.402...21N} and they are well fitted by an Einasto profile \citep{1965TrAlm...5...87E}, in particular for MW-like halos \citep{2017MNRAS.469.1101G}. As the theory of the formation of our Universe dominated by the DM is established, the $\Lambda$CDM model can now be confronted with observations. 

\subsection{The historical cusp-core problem}

\begin{figure}[!t]
\centering
\includegraphics[width=0.75\textwidth]{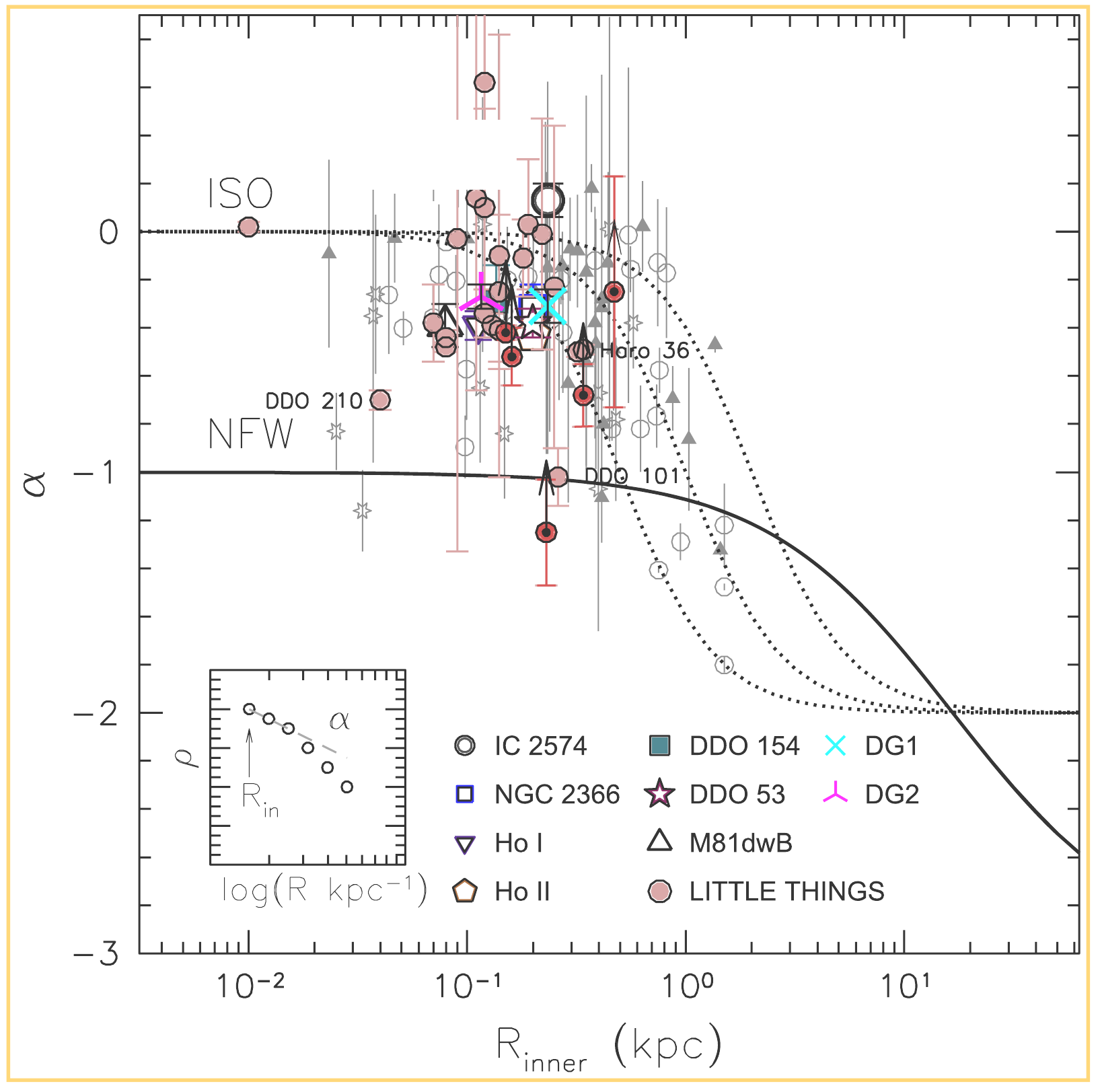}
\caption{{\it\bf Cusp-core problem:} Inner slope $\alpha$ of the density profiles as a function of the radius of the innermost point, within which $\alpha$ is measured. The theoretical slopes of a pseudo-isothermal halo are over-plotted with dotted lines for a core size of 0.5 (leftmost), 1 (centre), and 2 (rightmost) kpc. The solid line represents an NFW model \citep{1996ApJ...462..563N}. The pseudo-isothermal model is preferred over the NFW model to explain the observational data (see Equations~\eqref{eq1} and ~\eqref{eq2}). The figure is reprinted from \cite{2015AJ....149..180O}.}
\label{c1f1}
\end{figure}

\medskip

In order to find good observational probes of the DM distribution, it is essential that the dynamics of selected galaxies are dominated by DM, i.e with $M_{*}/M_{\rm DM}< 10^{-2}$. That is the reason why, the mass regime, which has been studied most extensively, is at the dwarf galaxy scale. "Dwarfs" usually refer to galaxies with $M_{*}<10^9$ M$_{\odot}$. About forty dwarf galaxies have been discovered in the Local Group, which encompasses our MW galaxy and Andromeda galaxy \citep{1998mcdg.proc...53M}. The dwarf population of the MW exhibits various different morphological types. Dwarfs can be divided into roughly two groups: those lack gas and have no ongoing star formation, correspond to dwarf spheroidals (dSphs) and those with gas and ongoing star formation are called dwarf Irregulars (dIrrs) \citep{2012AJ....144....4M}. 

\medskip

Early measurements of the HI rotation curves of gas-rich dwarf galaxies highlighted, for the first time, a large discrepancy between the observed rotation velocities and those predicted by $\Lambda$CDM simulations, especially in the inner parts \citep{1994Natur.370..629M,1994ApJ...427L...1F,1995ApJ...447L..25B}. The DM density profile of dIrrs can be inferred by measuring the rotation of either the gas with HI or using the stellar H$\alpha$ emission line. The rotation curve is derived from the observed line-of-sight velocity at any position in the galaxy velocity field. After the data analysis, the central DM distributions in these DM-dominated galaxies were found to be inconsistent with the $1/r$ behavior of cuspy profiles and indicate the presence of a constant-density core. These latter implies $\rho(r)\propto r^{-\alpha}$ with $\alpha=0$ in the inner regions. This discrepancy between observations and DM-only simulations led to the original small-scale problem, which has now become known as the cusp-core problem \citep{1994Natur.370..629M,1994ApJ...427L...1F}.

\medskip

However, it was argued that systematic effects could be responsible for the core signature in the rotation curve \citep{2002ApJ...575..801M,2003MNRAS.340..657D,2004ApJ...617.1059R}. The absence of a comprehensive and satisfactory resolution has also led to a wide range of different conclusions concerning the DM inner profile \citep{2001MNRAS.325.1017V}. As a consequence, the NFW form cannot be ruled out \citep{2005AJ....129.2119S}. Recent surveys of nearby dwarf galaxies, THINGS and LITTLE THINGS have offered ultra-high-resolution rotation curve data \citep{2008AJ....136.2563W,2012AJ....144..134H}. Indeed, high-resolution velocity fields were used to derive stronger constraints on the DM distributions in galaxies \citep{2008AJ....136.2720T,2008AJ....136.2761O,2008AJ....136.2648D}. A core profile represented by the pseudo-isothermal model is preferred over the NFW profile to explain the observational data (see Figure~\ref{c1f1}). The mass distribution of the pseudo-isothermal sphere is given by: 
\begin{equation}
    \rho(r)=\frac{\rho_0}{1+\left(r/R_{\mathrm{c}}\right)^2},
    \label{eq2}
\end{equation}
where $\rho_0$ and $R_{\mathrm{c}}$ are the central density and the core radius of the DM halo, respectively. By reaching the necessary resolution to alleviate some systematic effects, the logarithmic inner slope $\alpha$ of their DM halo densities were found to be about $\alpha=-0.32\pm0.24$ \citep{2011AJ....141..193O,2015AJ....149..180O}. Thus, these recent measurements of galaxy rotation curves in dIrrs reinforced the historical disagreement with the $\Lambda$CDM prediction at small scales (see Figure~\ref{c1f1}). 

\medskip

\subsection{A promising solution for gas-rich dwarf galaxies}

One of the key predictions of the $\Lambda$CDM paradigm is that DM assembles into halos that develop cuspy density profiles following the NFW form in the absence of baryonic effects. Indeed, the cusp-core problem in dIrrs was established without the inclusion of baryons. That is the reason why baryonic physics appeared as a natural solution within the $\Lambda$CDM framework. Moreover, the size of derived DM cores is typically on the order of a few kpcs, where baryons start to play an important role. Since DM interacts only gravitationally, baryons can affect it through the gravitational potential. The most promising solution, which was designated to explain this discrepancy at small scales, is stellar feedback \citep{1996MNRAS.283L..72N}. This feedback process consists of all interactions of stars with the interstellar medium, which is mostly filled with gas. Contrary to radiative and chemical feedback, this mechanical feedback acts as an energy injection of
massive stars in form of winds or SN explosions \citep{2005SSRv..116..625C,1991ApJ...379...52W,1978MNRAS.183..341W,1974MNRAS.169..229L,1986ApJ...303...39D}. Besides, it was established that at the dwarf scale, stellar feedback dominates over other feedback processes such as black hole feedback, as it mainly comes from high mass stars.

\begin{figure}[!t]
\centering
\includegraphics[width=0.75\textwidth]{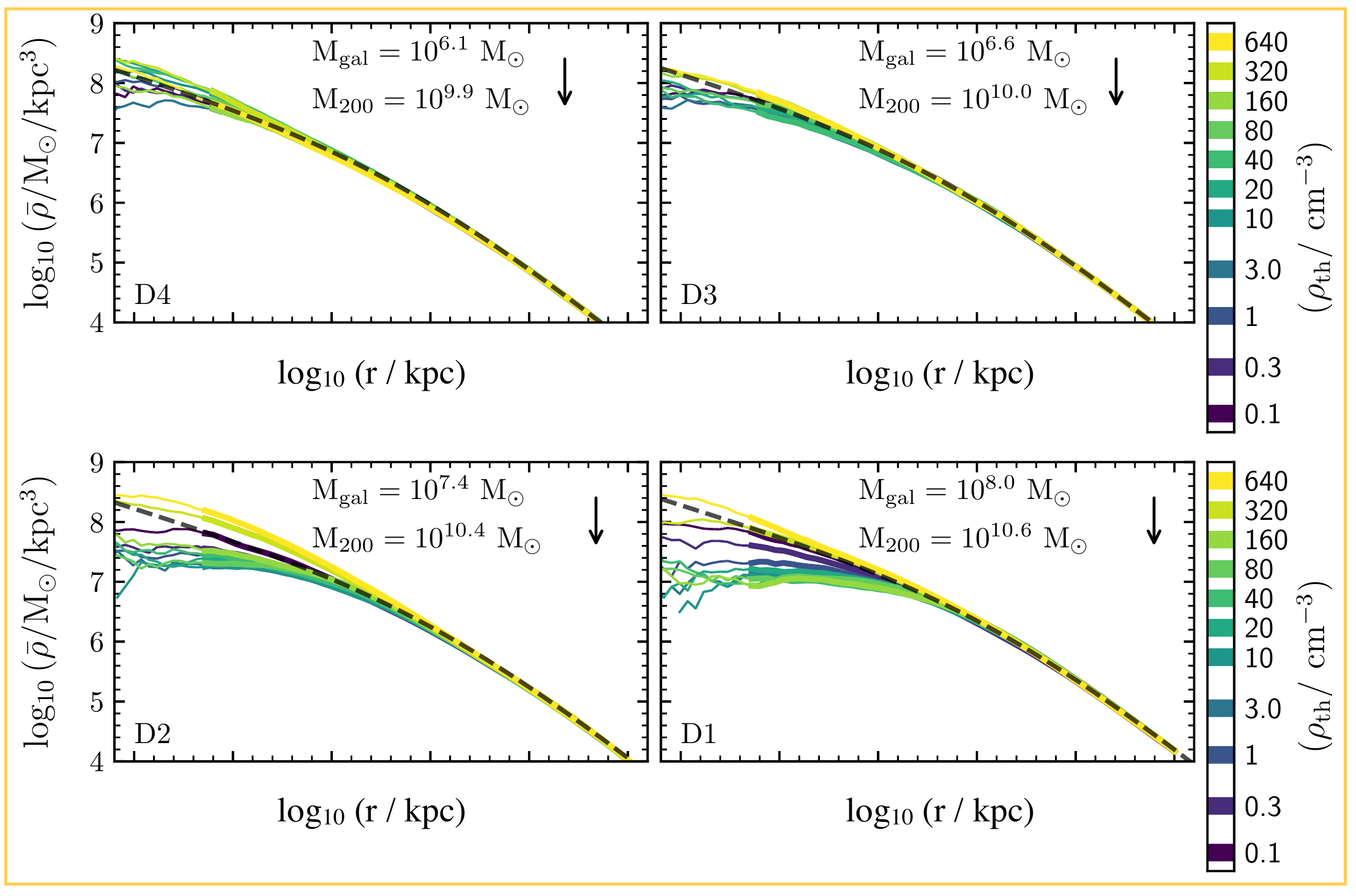}
\caption{{\it\bf Cusp-to-core transition:} Mean enclosed DM density profiles at $z=0$ of four dwarf galaxies with stellar masses between $10^6$ and $10^8$ M$_{\odot}$ for different gas density thresholds for star formation, $\rho_{\mathrm{th}}$, compared to the NFW profile (dashed curve) in numerical simulations that incorporate baryonic feedback. The virial radius of the halos is indicated by black arrows. The value of $\rho_{\mathrm{th}}$ varies from 0.1 to 640 cm$^{-3}$. For $\rho_{\mathrm{th}}=$ 0.1 cm$^{-3}$, the DM profiles are all consistent with NFW form above the convergence radius defined by \cite{2003MNRAS.338...14P}. This radius indicates the region within which numerical convergence is not achieved because of two-body relaxation. For higher values of $\rho_{\mathrm{th}}$, the density profiles depart systematically from NFW in some cases. The dependence of the core radius on the halo mass is highlighted over a wide range of the gas density threshold. Very low-mass dwarfs do not exhibit large DM cores as in earlier works \citep{2014MNRAS.441.2986D,2016MNRAS.456.3542T,2017ARA&A..55..343B}. The particular choice of $\rho_{\mathrm{th}}$ determines the size of the core. This figure is reprinted from \protect\cite{2019MNRAS.488.2387B}.}
\label{c1f5}
\end{figure}

\medskip

Even if baryons steepen the DM potential well when they cool and accumulate at their centre \citep{1986ApJ...301...27B,2004ApJ...616...16G,2010MNRAS.407..435A,2015MNRAS.453L..58S}, this feedback mechanism is able to alter the DM distribution by generating significant movements of the gas. Indeed, gas gathers in DM halos and feedback can expel large amounts of gas from the bottom of their potential well \citep{1996MNRAS.283L..72N,2002MNRAS.333..299G,2005MNRAS.356..107R,2014ApJ...786...87B,2008Sci...319..174M,2012ApJ...744L...9M,2012MNRAS.421.3464P,2014Natur.506..171P,2014ARA&A..52..415M,2020MNRAS.491.4523F,2013MNRAS.432.1947M,2017ApJ...839L..13S}. A fraction of this gas then cools and returns to the centre, generating repeated cycles of significant gas outflows which, in turn, cause rapid fluctuations of the gravitational potential. These potential fluctuations dynamically heat the DM and lead to the formation of a core. As a result, this baryonic process transforms a central DM cusp ($\alpha=$-1) into a core ($\alpha=$0). The gradual dispersion of the DM particles away from the centre of the halo is ultimately responsible for core creation. More precisely, these fluctuations in the potential transfer energy into DM particles and expand the DM distribution. Thus, one solution to the cusp-core problem in dIrrs is that a DM heating through stellar feedback generates a cusp-to-core transition for the DM halo within the CDM paradigm. 

\medskip

Cosmological hydrodynamical simulations performed with different codes such as GASOLINE \citep{2014MNRAS.441.2986D,2016MNRAS.456.3542T,2010Natur.463..203G,2012ApJ...761...71Z}, FIRE \citep{2018MNRAS.480..800H,2015MNRAS.454.2981C,2017MNRAS.471.3547F,2016ApJ...827L..23W,2015MNRAS.454.2092O,2017MNRAS.471.1709G}, RAMSES \citep{2017MNRAS.472.2153P} and GADGET \citep{2010MNRAS.402.1536S,2010MNRAS.405.2161D,2016MNRAS.457.1931S,2016MNRAS.457..844F,2019MNRAS.486.4790B,2019MNRAS.488.2387B} have proved the efficiency of these feedback mechanisms for core creation. Many of these most advanced hydrodynamic simulations with different feedback implementations are able to produce core-like density profiles as inferred from rotation curves such as those shown in Figure~\ref{c1f1}. Figure~\ref{c1f5}, reprinted from \cite{2019MNRAS.488.2387B} shows that for $\rho_{\mathrm{th}}=$ 0.1 cm$^{-3}$, the DM profiles are all consistent with NFW form above the convergence radius defined by \cite{2003MNRAS.338...14P}. For higher values of $\rho_{\mathrm{th}}$, the density profiles depart systematically from NFW in some cases. The dependence of the core radius on the halo mass is highlighted over a wide range of the gas density threshold. This confirms that very low-mass dwarfs do not exhibit large DM cores as in earlier works \citep{2014MNRAS.441.2986D,2016MNRAS.456.3542T,2017ARA&A..55..343B}. Moreover, it is also demonstrated that the particular choice of $\rho_{\mathrm{th}}$ determines the size of the core (see Figure~\ref{c1f5}). However, it was concluded that a value of $\rho_{\mathrm{th}}$ higher than the mean interstellar medium density is necessary for forming cores induced by stellar feedback
\citep{2019MNRAS.486..655D,2019MNRAS.488.2387B}. However, these simulations have shown that cores form efficiently only in a narrow range of stellar-halo mass, which corresponds to bright dwarf galaxies ($M_{*}=10^7-10^9$ M$_{\odot}$) (see Figure~\ref{c1f5}). It was also suggested that the inner slope of DM halos is mass-dependent \citep{2014MNRAS.441.2986D,2012MNRAS.421.3464P}. Indeed, a relationship was established between the slope $\alpha$ and the stellar-halo mass fraction, $M_*/M_{\mathrm{vir}}$, of simulated galaxies \citep{2014MNRAS.441.2986D,2016MNRAS.456.3542T,2018MNRAS.480..800H,2015MNRAS.454.2981C,2015MNRAS.454...83W}. As a result, there is a characteristic mass-ratio of $M_*/M_{\mathrm{vir}}=$0.005 for efficient core formation below which DM halos remain similar to the cuspy NFW profile predicted by DM-only simulations. In fact, DM halos become more cored as $M_*/M_{\mathrm{vir}}$ increases to this characteristic mass-ratio. On the contrary, it was demonstrated that it is possible to induce cusp-to-core transition in dwarfs of all stellar masses \citep{2016MNRAS.459.2573R}. This is made feasible by the fact that the stellar masses of dwarfs were slightly overestimated compared to those of cosmological simulations, such as Illustris TNG \citep{2018MNRAS.475..676S}. As this gives a good match to observations of dIrrs, it suggests questioning $M_*/M_{\mathrm{vir}}$ for dwarf galaxies in our cosmological models.
  
\medskip

Even if hydrodynamical simulations alleviate this $\Lambda$CDM tension by creating cores, its significance depends on the feedback model \citep{,2016MNRAS.457.1931S,2016MNRAS.457..844F,2019MNRAS.486.4790B,2019MNRAS.488.2387B,2015MNRAS.452.3650O}. Indeed, galaxies without a sufficient star formation are unlikely to have cores due to the lack of energy from feedback \citep{2012ApJ...759L..42P}. It was also argued that the timing of star formation relative to DM halo growth can also affect core formation. Cusps can regenerate from the core induced by the feedback as a result of DM rich mergers \citep{2015MNRAS.454.2092O}. As discussed, the gas density threshold is a crucial feedback parameter for producing cores in galaxies. Cosmological simulations with low-density thresholds for star formation such as APOSTLE, Auriga and GEAR \citep{2019MNRAS.486.4790B, 2018A&A...616A..96R} have been shown to not exhibit DM cores.

\subsection{Review plan}

This review aims to focus on gas-poor dwarf galaxies with $M_{*}=10^5-10^7$M$_{\odot}$ and the previous solution seems to be only designed for gas-rich dwarfs such as dIrrs, which are still forming stars today. In dSphs, star formation ceased shortly after the beginning of the Universe. In fact, they have characteristically old stellar populations and are generally devoid of gas. All hydrodynamical simulations find that baryonic feedback is negligible in the regime of classical dwarfs ($M_*/M_{\mathrm{vir}}<10^{-4}-10^{-3}$), as expected on energetic grounds \citep{2012ApJ...759L..42P,2013MNRAS.433.3539G}.
Thus, it seems more and more challenging to find solutions without invoking baryonic processes as DM cores tend to persist even for these dwarfs, which are rather DM-dominated. In the absence of new solutions in $\Lambda$CDM, it will inevitable to directly question the nature of the DM to reproduce the observations at small scales in these galaxies.

\medskip

Although the CDM paradigm can successfully explain various observations at different scales, this discrepancy at small scales remains one of the greatest challenges faced by this DM theory (see \cite{2010AdAst2010E...5D} for a detailed review on the observational challenges and see \cite{2018MNRAS.474.1398G,2017ARA&A..55..343B,2020Univ....6..107D,2018RvMP...90d5002B} for global reviews related to the cusp-core problem). Even if we focus only on the cusp-core problem in this review, there are other tensions of the $\Lambda$CDM model at small scales, which are the missing satellites problem, the too big to fail problem and the alignment of the substructures in the Galactic halo \citep{1999ApJ...522...82K,2012JCAP...05..030S,2011MNRAS.415L..40B,2017ARA&A..55..343B}. This review is intended to give an overview of the current observational and theoretical status concerning the DM distribution at small scales for the gas-poor dwarf spheroidal galaxies but also trying to give new directions to solve this challenging problem.

\section{The cusp-core problem in gas-poor Milky Way satellites}

Close to the MW and M31, one finds predominantly dwarf spheroidals. These dwarfs are among the most DM-dominated galaxies in the Universe \citep{2013NewAR..57...52B,2013pss5.book.1039W}. As DM constitutes 90$\%$ or more of the total mass in these dwarf spheroidals the dynamics are determined entirely by the gravitational field of the DM. Therefore, these systems provide an excellent laboratory to study DM distribution at small scales. The eight most common dwarf spheroidals are the galaxies orbiting around our Galaxy and also named "classical" dwarfs. These dSphs have a stellar component of about $10^6$ M$_{\odot}$ embedded in a DM halo of about $10^9$ M$_{\odot}$ (see Table~\ref{tab1}). As depicted by the Table, the DM masses are poorly constrained. Its estimate is limited to two observed values: the line-of-sight velocity dispersion and the projected half-light radius. Moreover, we underline that only one measurement of the line-of-sight velocity dispersion per galaxy is available for the MW dwarfs. 

\begin{table}
\begin{center}
\label{tab:landscape}
\begin{tabular}{cccccccccccc}
 \hline
 \\
Dwarf & M$_{*}$ &  M$_{\rm vir}$ & M$_*$/M$_{\mathrm{vir}}$ & Member stars \\
      & [$10^6$ M$_{\odot}$] & [$10^9$ M$_{\odot}$] & [$10^{-4}$] &  \\
    \hline
    \\
Fornax & 14 $\pm$ 4 & 2.5$^{+22}_{-1}$ & 56 & 2573 \\
Leo $I$ & 3.4 $\pm$ 1.1  & 25$^{+6}_{-0}$ & 1.3 & 328\\
Sculptor & 1.4 $\pm$ 0.6 & 25$^{+14}_{-20}$ & 0.5 & 1351\\
Leo II & 0.59 $\pm$ 0.18 & 25$^{+14}_{-24}$ & 0.23 & 186\\
Sextans & 0.41 $\pm$ 0.19 & 0.4$^{+0.39}_{-0.27}$ & 10.25 & 417\\
Carina & 0.24 $\pm$ 0.1 & 2.0$^{+37}_{-1.8}$ & 1.2 & 767\\
Ursa Minor & 0.20 $\pm$ 0.09 & 25$^{+14}_{-20}$ & 0.07 & 430\\
Draco & 0.27 $\pm$ 0.04 & 25$^{+14}_{-15}$ & 0.1 & 504\\
    \hline
\end{tabular}
\caption{{\it Classical dwarf spheroidal galaxies:} From left to right, the columns give for each gas-poor dwarfs: the galaxy stellar mass from \protect\cite{2018ApJ...860...76H} assuming a mass-to-light ratio of 1, the DM mass from \protect\cite{2018MNRAS.481.5073E}, the stellar-to-halo mass ratio and the the number of kinematic member stars from \protect\cite{2019MNRAS.484.1401R}. We have chosen to show the DM mass of the dwarfs assuming the presence of a core in order to underline that these systems are even more dominated by DM following this density profile. The ratios calculated here are only intended to give an idea of the scale in regard to the uncertainties on the DM masses.}
\label{tab1}
\end{center}
\end{table}

\subsection{Dynamical models}

As most dwarf galaxies are devoid of gas, it is necessary to look at the kinematics of their stars in order to probe their DM inner region. Indeed, rotation curve measurements are impossible for dSphs as they lack rotating gas components. However, only line-of-sight velocities of stars are observable. The line-of-sight velocity dispersion of these stars from the spherical Jeans equation \citep{1980MNRAS.190..873B,2008gady.book.....B} can be written as \citep{1982MNRAS.200..361B}:
\begin{equation}
    \sigma^2_{\mathrm{los}}=\frac{2}{\Sigma_*(R)}\int_{R}^{\infty}\left(1-\beta\frac{R^2}{r^2}\right)\frac{\nu(r)\sigma^2_{\mathrm{r}}(r)r}{\sqrt{r^2 - R^2}} \mathrm{d}r,
\end{equation}
where $\Sigma_*(R)$ is the surface mass density at projected radius R and the radial velocity dispersion $\sigma^2_{\mathrm{r}}(r)$ is defined as:
\begin{equation}
    \sigma^2_{\mathrm{r}}(r)=\frac{1}{\nu(r)g(r)}\int_{r}^{\infty}\frac{GM(u)\nu(u)}{u^2}g(u) \mathrm{d}u,
\end{equation}
with
\begin{equation}
    g(r)= \exp\left(2\int\frac{\beta(r)}{r}\right),
\end{equation}
where $\nu(r)$ and $\beta(r)$ are the radial density profile and the velocity anisotropy, which describes the orbital structure of the stellar system, respectively. $\beta=$ 0, 1 and $-\infty$ correspond to an isotropic, fully radial and fully tangential distributions, respectively. 
\begin{figure}[!t]
\centering
\includegraphics[width=0.75\textwidth]{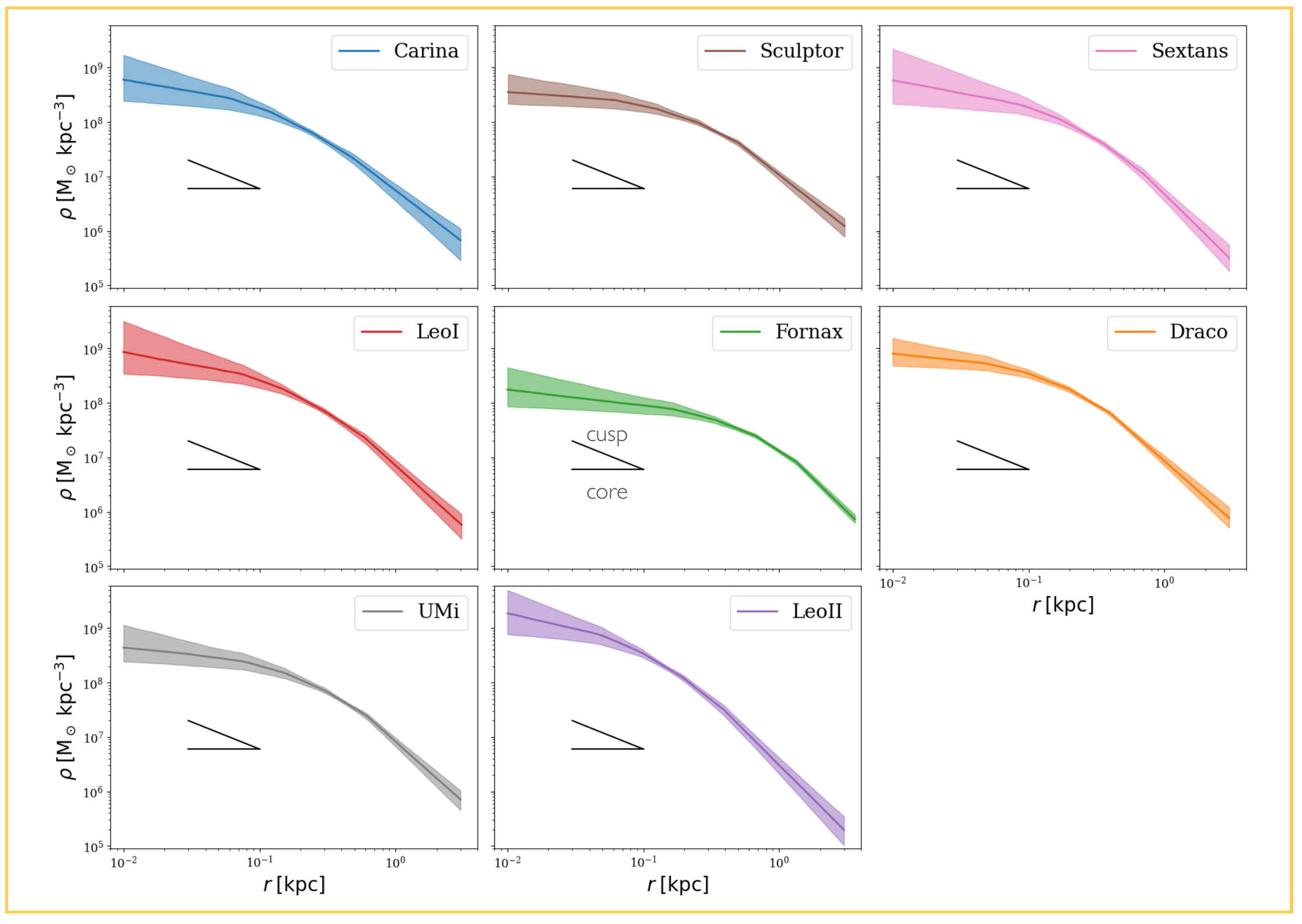}
\caption{{\it\bf Dynamical modeling:} DM density profiles of the eight MW classical dwarf galaxies derived from the stellar kinematics \citep{2019MNRAS.484.1401R}. The shaded regions mark the 68$\%$ confidence intervals of the model. In this interval, it is hard to distinguish between cusp and core for the MW satellites according to GravSphere fits.}
\label{c1f2}
\end{figure}

This technique allows the measurement of the central DM density profile in galaxies as the line-of-sight velocity dispersion of stars depends on the mass profile $M(r)$ \citep{2009ApJ...704.1274W,2017MNRAS.471.4541R,2019arXiv191109124G}. However, there is a degeneracy between the radial density profile of DM, $\rho(r)$, and the unknown orbit distribution of the stars. This latter is typically characterized by the velocity anisotropy parameter $\beta$ which is hard to constrain with only line-of-sight velocities \citep{1982MNRAS.200..361B,1990AJ.....99.1548M,2009MNRAS.393L..50E,2017MNRAS.471.4541R}. That is the reason why analyses of the line-of-sight velocities in dwarf galaxies have led to contradictory conclusions. Some authors conclude that the kinematic data require DM core \citep{2007ApJ...663..948G,2011ApJ...742...20W,2012ApJ...754L..39A}, while others found that the data are also consistent with the NFW form \citep{2010MNRAS.408.2364S,2013MNRAS.433.3173B,2014MNRAS.441.1584R}. For the brighter MW dwarfs, this degeneracy can be broken by using metallicity or colour to split the stars into distinct components \citep{2008ApJ...681L..13B,2011ApJ...742...20W,2012ApJ...754L..39A}. Other methods have been proposed to break this degeneracy by using higher-order velocity moments \citep{2009MNRAS.394L.102L}, Schwarzschild methods \citep{2013MNRAS.433.3173B,2013ApJ...763...91J}, and proper motions \citep{2002MNRAS.330..778W,2007ApJ...657L...1S,2018NatAs...2..156M}. Indeed, together with line-of-sight velocities and positions on the sky, stellar proper motions, which are the two additional transverse velocity components, provide five out of the six phase-space coordinates of the stars. The degeneracy may be also broken by including the fourth-order projected virial theorem \citep{1990AJ.....99.1548M}. A non-parametric Jeans method, namely GravSphere, employs the additional constraints from the virial shape parameters in their analysis \citep{2017MNRAS.471.4541R}. This higher-order Jeans analysis method has been shown to successfully recover DM density distributions of simulated dwarfs above half of the projected half-light radius \citep{2017MNRAS.471.4541R,2019arXiv191109124G}. Incorporating proper motions of stars was also employed to ameliorate this mass-anisotropy degeneracy \citep{2007ApJ...657L...1S,2020MNRAS.493.5825L}.

\medskip

In Figure~\ref{c1f2}, the DM density profile of eight dwarf spheroidal galaxies was estimated by using stellar kinematics \citep{2019MNRAS.484.1401R}. In the 68$\%$ confidence interval, it is hard to distinguish between cusp and core for the MW satellites according to GravSphere fits (see Figure~\ref{c1f2}). As their profile is better constrained at a radius of 150 pc, it was established that seven dwarfs have a central DM density $\rho$(150 pc) consistent with a cusp and only Fornax had a $\rho$(150 pc) consistent with a DM core \citep{2019MNRAS.484.1401R}. However, it still leaves room for DM cores of less than 100 pc based on GravSphere model uncertainties (see Figure~\ref{c1f2}). Accounting for dwarfs in dynamical equilibrium, \cite{2020ApJ...904...45H} found also a diversity of DM density profiles with many actually favoring cuspy profiles. 

\medskip

Besides, the Jeans method usually assumes dwarfs as spherical systems for simplicity. However, it was claimed that the stellar component of the dwarfs is actually non-spherical \citep{1995MNRAS.277.1354I,2012AJ....144....4M}. As they formed in a hierarchical manner, DM halos are also expected to be non-spherical \citep{2014MNRAS.439.2863V,2007ApJ...671.1135K,2002ApJ...574..538J}. \cite{2015ApJ...810...22H} applied the Jeans technique to line-of-sight velocity dispersion profiles of seven MW dwarfs. Contrary to \cite{2019MNRAS.484.1401R}, they found that five dwarfs including Fornax have a cored central density profile \citep{2015ApJ...810...22H}. Indeed, non-spherical halo models seem to reveal a more diffuse DM distribution in the inner region of dwarfs. By using the Jeans modelling, it is generally assumed that the MW tides have not had much impact on the stellar kinematics of dwarfs. However, there is a different physical effect, which is not due to tidal stripping and occurs only for highly eccentric orbits, namely tidal shocking \citep{1999ApJ...513..626G,1999ApJ...522..935G,1987degc.book.....S}. In fact, the MW tidal shocks can bring sufficient kinetic energy to heavily affect the velocity dispersions of stars. Since DM calculations are based on stellar kinematic measurements, one may wonder whether they could have been corrupted by the fact that dSphs were out of equilibrium because of MW tides \citep{2019ApJ...883..171H,2018ApJ...860...76H,2020ApJ...892....3H}. It was demonstrated that it takes more than three dynamical times for a system to virialize after a perturbation \citep{1999ApJ...514..109G}. Finally, this questions the validity of the dynamical mass estimate using the Jeans equation, hence on the estimates of the DM amount in MW dwarfs. 

\medskip

Moreover, a recent study highlights the need for a large number of kinematic member stars for dwarfs in order to accurately determine the DM inner profile \citep{2021MNRAS.507.4715C}. By using mock observations, they showed that it is necessary to measure about 10000 stars within a single dwarf galaxy to infer correctly the DM distribution at small scales. With data sets of fewer than 10000 stars, it appears that the DM density distribution is biased towards a steeper inner profile than the true distribution by applying the Jeans method. This effect could explain why \cite{2019MNRAS.484.1401R,2020ApJ...904...45H} found that the majority of classical dwarfs exhibit cuspy profiles. As described in Table~\ref{tab1}, the number of stars used by \cite{2019MNRAS.484.1401R} to infer DM profiles of MW dwarfs is well below what is recommended by \cite{2021MNRAS.507.4715C}. 

\begin{figure}[!t]
\centering
\includegraphics[width=0.75\textwidth]{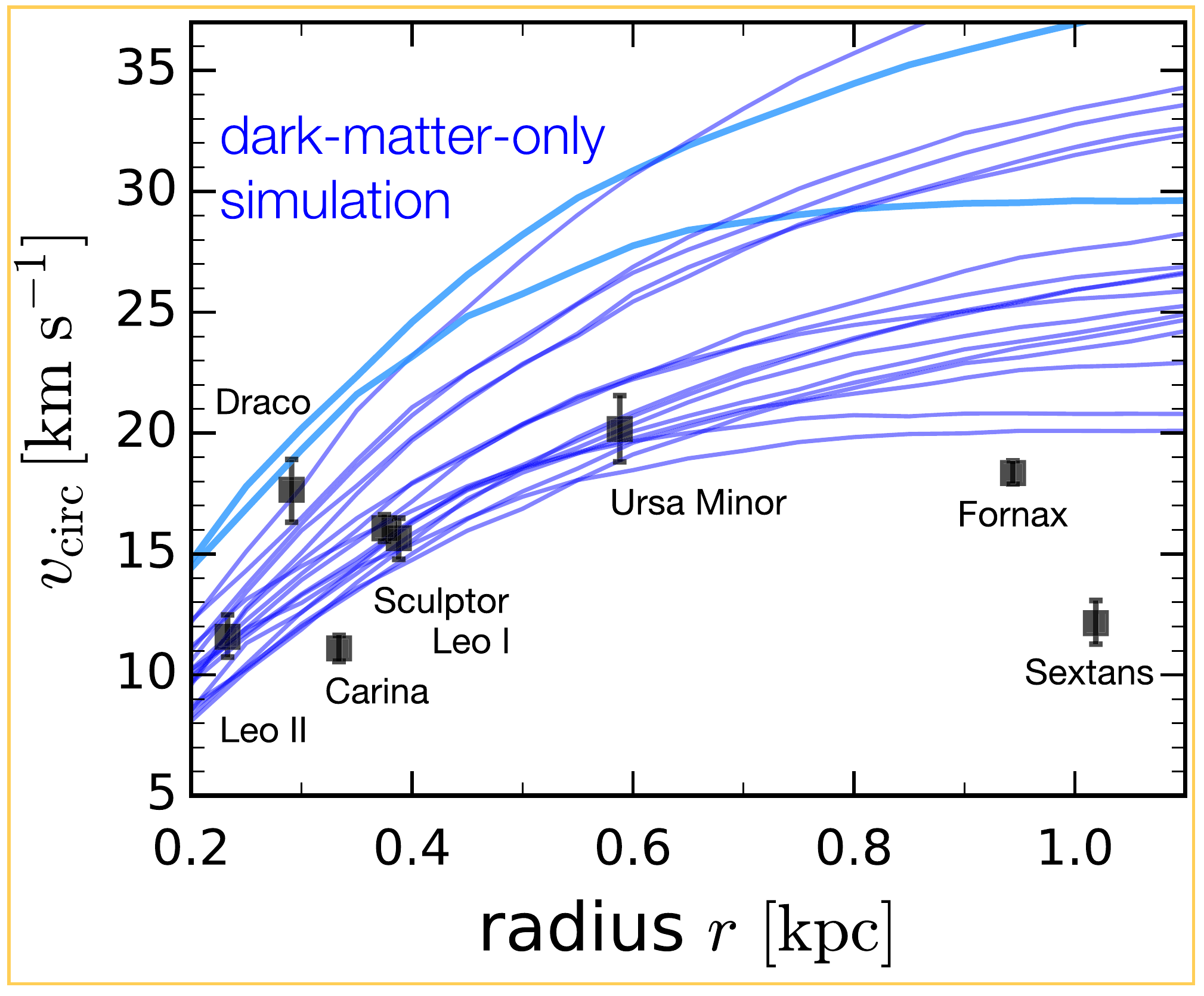}
\caption{{\it\bf Rotation curves of classical dwarfs:} Circular velocity of 19 subhalos in the dark-matter-only simulation at $z=0$. Black points show observed gas-poor satellites of the MW \protect\citep{2010MNRAS.406.1220W}. Only five subhalos from the GIZMO cosmological simulation are consistent with Ursa Minor, Draco, Sculptor, Leo I, and Leo II. This figure is reprinted from \protect\cite{2016ApJ...827L..23W}.}
\label{c1f4}
\end{figure}

\medskip

Despite the complexity of the Jeans analysis, dynamical models are often claimed to require shallower density profile slopes that are consistent with a core at their centre \citep{2011ApJ...742...20W,2012ApJ...754L..39A,2012MNRAS.419..184A,2014ApJ...789...63A,2019MNRAS.484.1401R}. The mass measurements of \cite{2010MNRAS.406.1220W} reinforce the prediction for the presence of DM cores in gas-poor MW satellites. Figure~\ref{c1f4}, reprinted from \cite{2016ApJ...827L..23W}, shows the circular velocity of 19 subhalos in the DM-only simulation at $z=0$. Only five subhalos from the GIZMO cosmological simulation are consistent with Ursa Minor, Draco, Sculptor, Leo I, and Leo II. As a result, the other DM subhalos are too dense. One way to reduce the inner DM density of halos is core formation.

\subsection{Controversy on producing DM cores via supernova feedback}

Historically, stellar feedback was invoked to solve this issue. However, as described by Table~\ref{tab1}, the classical dwarfs, except Fornax, are in the mass ratio regime ($M_*/M_{\mathrm{vir}}<10^{-4}-10^{-3}$), where the contribution of feedback for creating DM core is negligible \citep{2012ApJ...759L..42P,2013MNRAS.433.3539G,2017MNRAS.472.2356M,2016MNRAS.456.3542T}. Besides, only Fornax has an extended star formation compared to the other dwarfs \citep{2019MNRAS.484.1401R}. That is the reason why stellar feedback via supernovae explosions could explain the presence of a DM core as star formation proceeds for long enough in Fornax \citep{2019MNRAS.484.1401R}. Nevertheless, star formation shut down 1.75 Gyr ago in Fornax and it was demonstrated that in $10^9$ M$_{\odot}$ dwarfs such as Fornax, DM cores induced by multiple repeated bursts needs $\sim$ 14 Gyr to fully form \citep{2016MNRAS.459.2573R}. The efficiency of this core creation mechanism in dwarf galaxies remains an intensive debate in the literature. We detailed the possible reasons for the disagreement between isolated and cosmological hydrodynamic simulations. 

\medskip

By using idealized high-resolution simulations, \cite{2016MNRAS.459.2573R} argued that DM cores form if star formation proceeds for long enough but this gives a stellar-to-halo mass ratios, which are not obtained CDM cosmological simulations in the dwarf regime \citep{2014MNRAS.441.2986D,2016MNRAS.456.3542T,2015MNRAS.454.2981C,2017MNRAS.472.2945R,2017MNRAS.472.2356M,2019MNRAS.486.4790B}. It was also reported that in the non-cosmological simulation of \cite{2016MNRAS.459.2573R}, some missing ingredients such as the UV background and the halo growth via mergers could be the most important sources of the differences. Indeed, the strong ionizing UV-background radiation has been identified as being capable of evaporating most of the gas in dwarf galaxies \citep{2014MNRAS.444..503N,2000ApJ...539..517B,1992MNRAS.256P..43E,1996MNRAS.278L..49Q}. In addition to the fact that cosmological simulations of galaxy formation tend to be more realistic than isolated simulations, the ingredients of galaxy formation are calibrated to the resulting structural properties of observed massive galaxies. As the efficiency of this cusp-to-core mechanism at dwarf scale is mainly determined by the gas density threshold for star formation, as discussed before, core formation thus depends on the baryon physics implemented in the simulation \citep{2019MNRAS.488.2387B,2019MNRAS.486..655D}. However, in some studies that have claimed the absence of DM cores in dwarfs, it was stressed that they are unable to resolve the clumpy interstellar medium, which is crucial for observing cusp-core transformations via supernova feedback \citep{2012MNRAS.421.3464P,2019MNRAS.484.1401R}. Recently, \cite{2021arXiv210301231B} pointed out that supernova feedback is a feasible mechanism of cusp-core transformation in dwarfs only if the supernova energy injection is longer than the dynamical timescale of DM particles in the inner halo. Moreover, they also stressed that DM heating is more efficient if baryons are more concentrated towards the centre of the galaxy \citep{2021arXiv210301231B}. Previous hydrodynamical simulations established a seeming connection between the presence of DM cores and the star formation history
of dwarf galaxies \citep{2016MNRAS.459.2573R} but there is a consensus that finding signatures of stellar feedback is not a sufficient condition for dwarfs to exhibit cores \citep{2021arXiv210301231B}. Indeed, baryon-induced cores in dwarfs would be difficult to distinguish from DM cores produced by other mechanisms such as in alternative DM theories \citep{2014MNRAS.444.3684V,2015AAS...22540205F,2019MNRAS.485.1008B}.

\medskip

Cosmological simulations have to cover a wide range of spatial and time scales. It is challenging to capture all relevant scales for this sort of simulation. That is the reason why dwarf galaxies continue to be one of the few areas where the CDM cosmological model has difficulties matching observations. IllustrisTNG as one of the most recent cosmological hydrodynamical simulation has achieved a mass and a spatial resolution of $\sim10^6$ M$_{\odot}$ and 0.2 kpc, respectively \citep{2018MNRAS.475..676S}. This main limitation has been pointed out as the source of the inconsistencies between predictions made by the CDM paradigm and observations. This also contribute to explain why the implementation of star formation and feedback is challenging. Besides it is imperative to remind that the problem of star formation is still unsolved at all redshifts and totally unconstrained at high redshift. Even non-cosmological simulations require making a number of choices and assumptions about the initial conditions as well as the input physics, they are essential to investigate the small-scale physics in dwarf galaxies and test various mechanisms. Idealized simulations have to be doing their bit. Thanks to these zoom simulations, we can achieve spatial resolutions up to 0.03 kpc with the VELA hydrodynamical simulation \citep{2014MNRAS.442.1545C} and mass resolution up to $6.2\times10^2$ M$_{\odot}$ with the NIHAO hydrodynamical simulation \citep{2015MNRAS.454...83W}. However, this limitation should soon be overcome by extreme resolution simulations \citep{2019MNRAS.490.4447W} allowing us to probe smaller physical scales than previously possible in cosmological simulations. These simulations with a mass and spatial resolution of 30 M$_{\odot}$ and $\sim0.1-0.4$ pc predict that the stellar do not significantly alter the density profile from cuspy to cored distribution \citep{2015MNRAS.454...83W}. This result is consistent with some of the lower resolution cosmological simulations \citep{2012ApJ...759L..42P,2015MNRAS.454.2092O}. However, feedback still needs to be modelled properly at these resolved scales.

\section{Solutions}

In this section, we investigate some of the most popular and promising solutions to the cusp-core problem. We are particularly interested in solutions, which could replace stellar feedback. Indeed, this core mechanism seems inefficient for most dSphs such as the gas-poor MW satellites. There are two main approaches that could solve this discrepancy between $\Lambda$CDM and observations. Cosmological solutions invoke a different spectrum at small scales \citep{2003ApJ...598...49Z}, different nature for DM particles, such as fuzzy and self-interacting DM \citep{2000ApJ...542..622C,2000NewA....5..103G,2000PhRvL..85.1158H,2000PhRvL..85.3335K,2000ApJ...534L.127P,2001ApJ...551..608S}, modified gravity theories \citep{1970MNRAS.150....1B,2020NewA...8001399G,2009PhRvD..79l4019B,2010PhRvD..81l7301L,2011JCAP...01..009D,2011JCAP...03..002Z,2016MNRAS.463.1944H,2020MNRAS.493.2373D,2018MNRAS.477.4187H} or Modified Newtonian dynamics \citep{1983ApJ...270..365M,2012LRR....15...10F,2008MNRAS.387.1481A,2014MNRAS.440..746A}. On the contrary, astrophysical solutions invoke sub-galactic baryonic physics within the $\Lambda$CDM paradigm such as stellar feedback \citep{1996MNRAS.283L..72N}. A common aspect of these two broad categories of solutions is that core creation has been identified as their main mechanism. In this review, we adopt a different classification. Our first category includes scenarios where DM cores emerge due to the flattening of initial $\Lambda$CDM cusps, named {\it cusp-to-core solutions}. We then focus on solutions which depict the natural emergence of DM cores such as in fuzzy and self-interacting DM theories, named {\it inherent core solutions}. 

\subsection{Cusps to cores}

It is admitted that a massive particle moving through an infinite, homogeneous and isotropic background of lighter particles experiences a force of dynamical friction given by 
\begin{equation}
    F(x,v)=2\pi G^2 \rho(x) \ln(1+\Lambda^2)\left(\mathrm{erf(X)}-\frac{2X}{\sqrt{\pi}}\exp(-X^2)\right)\frac{v}{\vert v \vert^3}M, 
\label{edyn}
\end{equation}
where this massive particle of mass $M$ at position $x$ is moving at velocity $v$ through a background density $\rho$ \citep{1943ApJ....97..255C}. The quantity $X$ is defined as $\vert v \vert /\sqrt(2)\sigma_{\mathrm{r}}$ with $\sigma_\mathrm{r}$ being the radial dispersion of lighter particles. The factor $\Lambda$ that goes into the Coulomb logarithm is taken to be
\begin{equation}
    \Lambda=\frac{r/\gamma}{\max(r_{\mathrm{hm}}, GM/\vert v \vert^2)},
\end{equation}
where $r_{\mathrm{hm}}$ is the half-mass radius of the massive particle and $\gamma$ is the absolute value of the logarithmic slope of the density, i.e. $\gamma=\vert d\ln(\rho)/d \ln(r) \vert$ \citep{2016MNRAS.463..858P}. The background medium composed of lighter particles produces an overdensity region behind it due to this friction between particles. The dynamical friction is responsible for a momentum loss by the massive object due to its gravitational interaction with its own gravitationally induced wake. The surrounding background medium, which consists of a combination of collisionless matter such as DM, is heated at an equal and opposite rate to the energy lost by the massive object. The rate of energy loss by the massive object is given by \cite{2001ApJ...560..636E}:
\begin{equation}
    \frac{\mathrm{dE}}{\mathrm{d}t}=M\frac{\mathrm{d}v}{\mathrm{d}t}v.
\end{equation}
An energy exchange occurs, increasing that of the medium particles at the expense of the perturber. If the perturber passes close to the central region of a dwarf galaxy, it could modify the DM inner structure via dynamical friction. During the perturber infall within the galaxy, it transfers part of its kinetic energy to the DM background through dynamical friction causing the DM particles in the central region of dwarfs to migrate outwards. The particle heating and migration in the central region of the galaxy is expected to lead to the flattening of the DM density profile. Indeed, at kpc scales, this collective effect induces potential fluctuations, which erode the central density cusp of the DM halo.

\subsubsection{Mergers with dwarf galaxies}

In our cosmological model, galaxies form in a hierarchical manner. They are formed on the one hand by mergers of pre-existing galaxies. High resolution $N$-body simulations have shown that as the satellite falls onto M31, it is slowed down by dynamical friction and its energy is transferred to the host halo. In this process, the initial cusp shallows down for over almost a decade and is well-fitted by a core profile \citep{2021ApJ...919...86B}. The efficiency of this mechanism depends on the mass, as depicted by the Equation~\eqref{edyn}, and on the orbit of the perturber. Indeed, it was suggested that the cusp-to-core transition occurs where the mass of the perturber within its tidal radius $r_{\rm t}$ roughly matches the enclosed mass of the DM background as follows:
\begin{equation}
    M_{\rm pert}(r_{\rm t})\sim M_{\rm DM}(r_{\rm p}),
\label{eread}
\end{equation}
where $r_{\rm p}$ is the instantaneous orbital radius of the perturber \citep{2010ApJ...725.1707G,2006MNRAS.366..429R}. Besides, it has been reported that merger events in which satellites fall on highly eccentric orbits onto their host halos can initiate core formation in a $\Lambda$CDM Universe where halos have cuspy profiles \citep{2021ApJ...919...86B}. In order to alter the DM distribution by scattering particles away from the centre, the satellite needs to pass through the central region of the galaxy. This condition is only satisfied with nearly radial orbits. 

\medskip

Now the question is to determine if cusps of the dwarf galaxies could be disrupted during mergers. Major mergers of dwarf galaxies are very rare after $z\sim$ 3 \citep{2018MNRAS.479..319F}. However, the CDM paradigm predicts that a very large number of DM substructures exist inside galactic halos \citep{2008Natur.454..735D,2008MNRAS.391.1685S}. Recently, Gaia DR2 data has provided additional evidence for these substructures \citep{2021MNRAS.502.2364B}. DM halos are growing with time notably by accretion of smaller halos, considered as DM subhalos. They interact gravitationally with all the components of the galaxy before becoming remnants of disrupted halos \citep{2019Galax...7...81Z}. It was pointed out that $10^9$ ($10^{10}$) M$_{\odot}$ dwarf halos have accreted 10-11 (13-14) subhalos with a mass ratio $10<M_{\mathrm{host}}/M_{\mathrm{sub}}<100$ over their history \citep{2020MNRAS.495L..12B}. This can be seen as minor mergers with these subhalos. Moreover, based on the approximated orbital distributions of satellites by \cite{2011MNRAS.412...49W}, it was shown that subhalos exhibit orbits, which are nearly radial with an eccentricity $e=0.85$ ($e=0.88$) at $z=3$ for $10^9$ ($10^{10}$) M$_{\odot}$ dwarf halos \citep{2020MNRAS.495L..12B}. Thus, dynamical perturbations induced by subhalo crossings, causing black holes (BHs) to vacate the galaxy centre, could also modify the spatial distribution of DM particles \citep{2020MNRAS.495L..12B}. Nevertheless, subhalos possess a very diffuse DM distribution. That is the reason why the condition described by Equation~\eqref{eread} is going to be satisfied only for small radii. \cite{2020MNRAS.495L..12B} demonstrated that the maximum offset reached by the BH due to heating from subhalos is 134 pc, which delimits the region where the DM distribution could have been significantly perturbed by subhalos. It was recently shown that in ultra-faint dwarf galaxies, the potential fluctuations could be also due to subhalo crossings but the DM distribution remains cuspy while it was flattened \citep{2021MNRAS.504.3509O}. Maybe the combination of stellar feedback and subhalos could then enhance the flattening of the central DM density \citep{2021MNRAS.504.3509O}. This mechanism seems unfortunately inefficient, particularly in the case of Fornax dwarf, which requires a DM core of size $\sim$ 1 kpc (see Figure~\ref{c1f2}). Recently, it was suggested that a dwarf major merger is needed to recover the current spatial distribution of globular clusters (GCs) in Fornax \citep{2020MNRAS.493..320L}. This ancient merger ($\sim$ 10 Gyr ago) could have contributed to the formation of the large DM core in Fornax dwarf.

\medskip

Besides, it was claimed that the stellar component of the satellite play a major role in core formation. Indeed, as this component is more concentrated compared to the DM of the satellite ($a_{*}/a_{\rm DM}=$ 0.1), it will further slow down the satellite during its infall and thus disturb the central region of the host even more prominently \citep{2021ApJ...919...86B}. As the stellar component of the satellite enhances the destruction of the cusp, galaxies with a low halo-to-stellar ratio could be promising candidates for such minor mergers but they are only found at very high redshift.

\subsubsection{Globular clusters and gas clumps}

\begin{figure}[!t]
\centering
\includegraphics[width=0.75\textwidth]{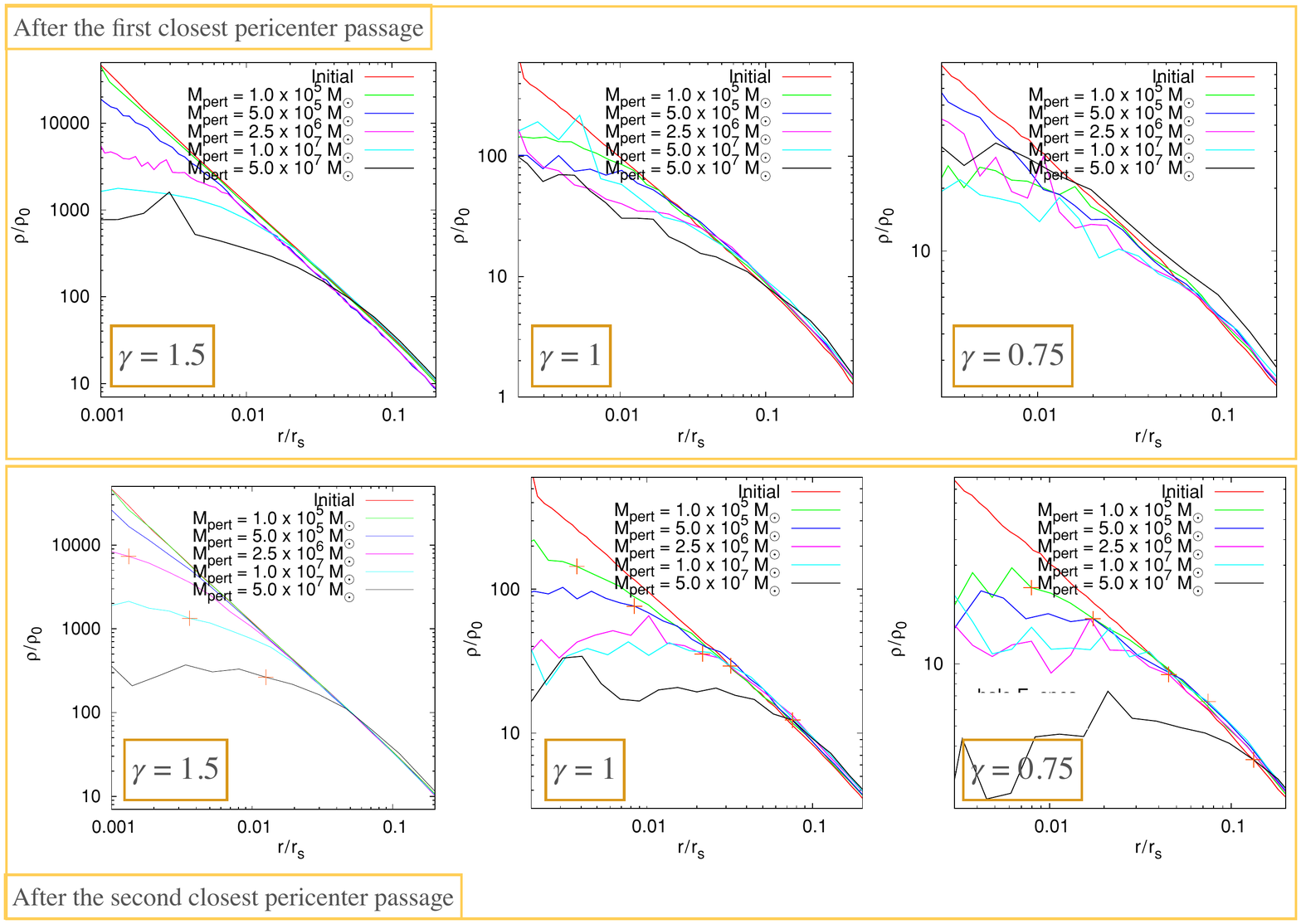}
\caption{{\it\bf Sinking of massive objects:} DM density profiles of the host halo after the first (upper panel) and second (lower panel) closest pericentre passage for the different perturber masses $M_{\mathrm{per}}=[10^5, 5\times10^5, 2.5\times10^6, 10^7, 5\times10^7]$. From left to right, the panels show halos with different initial absolute values of the logarithmic slope of the density $\gamma=\vert \mathrm{d}\ln(\rho)/\mathrm{d} \ln(r) \vert$. The DM density distribution changes significantly from cuspy to having a core. This figure is adapted from \cite{2010ApJ...725.1707G}.}
\label{c1f8}
\end{figure}

Galaxies also grow by accretion of a variety of objects such as GCs and gas clumps. That is the reason why such massive objects have also been proposed for transforming cusps into cores via the heating by dynamical friction \citep{2001ApJ...560..636E,2010ApJ...725.1707G,2015MNRAS.446.1820N,2011MNRAS.416.1118C,2014JCAP...12..051D,2011MNRAS.418.2527I}. Figure~\ref{c1f8}, adapted from \cite{2010ApJ...725.1707G}, shows the modification of the inner DM structure after the first (upper panel) and second (lower panel) closest pericentre passage of massive objects such as gas clumps or GCs with different masses. The perturbers were started within the cusp region. All simulations are shown using circular orbits for the infalling objects. We note that the response of different central cusps to sinking perturbers with a range of masses using $N$-body simulations occurs rapidly. The DM density distribution changes significantly from cuspy to having a core. Indeed, larger perturber masses lead to larger constant density central regions, as predicted by Equation~\eqref{eread}.

\medskip

The results of these works clearly indicated, as a proof of concept, that dynamical friction heating can have an important role in DM halos on different scales and the relevance of this process depends on the properties of the massive objects and of the host halo. Such a mechanism still requires another process to then destroy the gas clumps and GCs at the centre of the DM halo. Otherwise, the resulting inner stellar density would be too high to be consistent with observations \citep{2015MNRAS.446.1820N}. For the gas clumps, stellar feedback could dissolve these clumps. However, GCs form nuclear star clusters at the centre of galaxies but observations claim that none of the classical dwarfs exhibit a stellar nucleus at their centre.

\subsubsection{Globular clusters embedded in dark matter}

GCs are gravitationally bound groupings of mainly old stars, formed in the early phases of galaxy formation. Classically, it has been claimed that GCs do not contain DM because their dispersion velocities are too small. However, these measures are done at the centre of GCs, where the influence of DM is very small. Absence of evidence is not evidence for absence. Currently, there is no clear consensus on the formation of GCs, a subject which is hotly debated and which brings unique constraints on the formation of small-scale halos in the $\Lambda$CDM paradigm of galaxy formation. It has been proposed that GCs may have a galactic origin, where GCs are formed within DM minihalos in the early Universe \citep{1984ApJ...277..470P,2002ApJ...566L...1B,2005ApJ...619..258M,2016ApJ...831..204R}. Then, these GCs could have merged to become, later, a part of the present-day host galaxy. Until now, these DM halos have not been detected. More precisely, it was pointed out that the ratio of the mass in DM to stars in several GCs is less than unity \citep{2011ApJ...741...72C,2013MNRAS.428.3648I,2013JKAS...46..173S,1996ApJ...461L..13M,2009MNRAS.396.2051B,2010MNRAS.406.2732L,2015PhRvD..91j3514H}. Even if GCs are proven not to have a significant amount of DM today, it does not preclude them having been formed originally within a DM minihalo. A natural explanation is that they have lost their DM over time. Indeed, there are several internal dynamical processes which could eject DM from GCs such as DM decay \citep{2010PhRvD..81j3501P} and feedback processes \citep{2012MNRAS.421.3464P,2014MNRAS.443..985D}. It was also shown that GCs orbiting in the inner regions of their host galaxies may lose a large fraction of their primordial DM minihalos due to tidal stripping \citep{2006ApJ...640...22S,2005ApJ...619..258M,2012MNRAS.419.2063B,2002ApJ...566L...1B}. That is the reason why the main mechanism by which most GCs could have lost their DM minihalo is through severe tidal interactions with our Galaxy given their current positions. Nevertheless, GCs at a large distance from the MW centre could have retained a significant fraction of DM because it was not completely stripped by the Galaxy. Even if observations of these GCs such as NGC 2419 and MGC1 highlight that they do not possess significant DM today \citep{2011ApJ...741...72C,2013MNRAS.428.3648I}, it does not exclude the existence of DM in GCs but suggest that there is not necessarily a unique formation mechanism for GCs. 

\medskip

The motion of GCs embedded in DM minihalos inside the CDM halo of Fornax was studied by considering both early and recent accretion scenarios of GCs by Fornax with the most prevalent initial conditions taken from Illustris TNG-100 cosmological simulations \citep{2018MNRAS.473.4077P}. Using high-resolution simulations, these minor mergers involved perturbers with a low halo-to-stellar ratio ($\sim$10-20), which make GCs more massive. That is why they fall more rapidly towards the galaxy center \citep{2020MNRAS.492.3169B}. As expected, GC crossings near the Fornax centre induce a cusp-to-core transition of the DM halo. Moreover, if the five GCs were accreted recently, less than 3 Gyr ago, by Fornax, they should still be in orbit and no star cluster should form in the centre of Fornax in accordance with observations \citep{2020MNRAS.492.3169B}. By designing initial conditions such as GC orbits outside the Fornax tidal radius, avoiding the formation of a nuclear star cluster at the Fornax centre is possible without invoking this new dark component \citep{2009MNRAS.396..887A}. Nevertheless, crossings of GCs with a DM minihalo near the Fornax centre induce a cusp-to-core transition of the DM halo and hence resolve the cusp-core problem in this dwarf galaxy. The DM core size depends strongly on the frequency of GC crossings \citep{2020MNRAS.492.3169B}. It was subsequently highlighted that an infalling GC with a DM minihalo enhances core formation without forming a nuclear star cluster at the Fornax centre. Moreover, their results are in good agreement with the constraints on the DM mass of GCs as these clusters lost a large fraction of their DM minihalos. All of these aspects provide circumstantial evidence for the existence of DM halos in GCs. Nevertheless, it was pointed out that it should be regarded this as unlikely since GCs do not appear to be ubiquitous in local dwarf galaxies \citep{2020MNRAS.492.3169B}.

\subsubsection{Tidal interactions}

It is well-known that the classical dSph galaxies are satellites of the MW. Studies about dwarfs of the Local Group have revealed that DM cores can be generated through tidal stripping \citep{2018MNRAS.481.5073E,2018MNRAS.478.3879S,2019MNRAS.488.2743T}. By removing more and more bound particles, in an outside-in fashion, the effect of tides was also proposed as a solution to the cusp-core problem in a CDM universe. Indeed, the mass removal due to tidal forces can reduce the DM content at all scales even in the central region \citep{2017MNRAS.472.3378F}. This alternative mechanism was tested by \cite{2020arXiv201109482G} in order to explain the low inferred density in Fornax (see Figure~\ref{c1f2}). The majority of their Fornax analogues are able to lose DM from the inner 1 kpc due to tidal effects \citep{2020arXiv201109482G}. Even if this mechanism leads to a reduction in the DM density at all radii, the inner DM region of Fornax remains cuspy. \cite{2017MNRAS.472.3378F} have also stressed that there is a steepening of the central slope of the DM profile during satellite accretion by a MW-like galaxy. Even if dwarfs, which have shallower DM profiles due to feedback heating before accretion, evolve into cuspy DM halos \citep{2017MNRAS.472.3378F}.Thus, due to a low orbital pericentre in the MW or due to tidal interactions with other galaxies prior to infall, Fornax could exhibit a cuspy DM halo with its low density owing entirely to tides (see Figure~\ref{c1f2}). However, the absence of a DM core profile is still in tension with the kinematics of Fornax below its half-light radius, depicted in our Table~\ref{tab1} \citep{2019MNRAS.484.1401R}. Furthermore, there is currently no sign of tidal stripping in Fornax. In other words, no stream of unbound stars has not yet been detected \citep{2006AJ....131.2114W,2019ApJ...881..118W}.

\subsubsection{Cusp regeneration}

\begin{figure}[!t]
\centering
\includegraphics[width=0.75\textwidth]{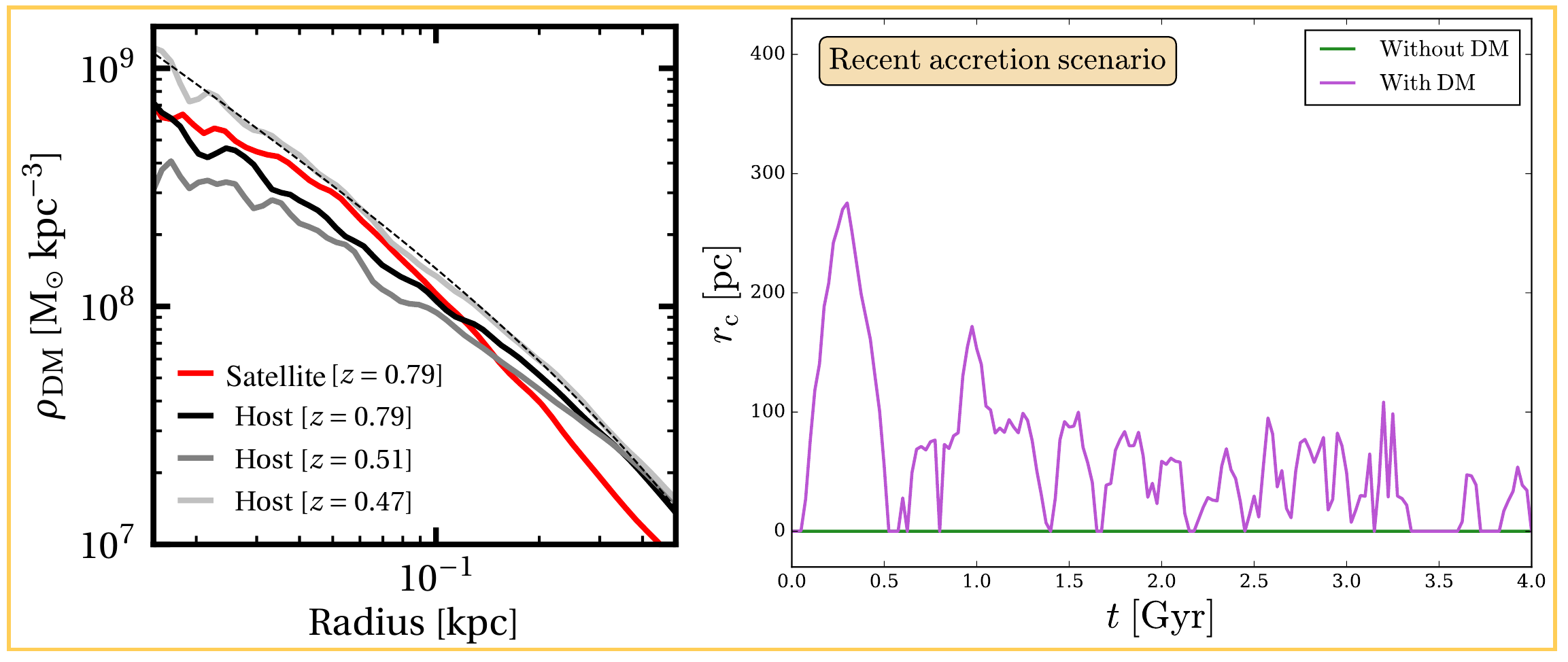}
\caption{{\it\bf Cusp regeneration in dwarfs:} {\it Left panel}: DM density profiles of a cuspy satellite (red line) and of the host halo at different redshifts (black, dark grey and light grey lines). A black dashed line represents a NFW profile fit to the host halo at $z = 0.79$. The merger with the satellite reforms the erased cusp. {\it Right panel}: Fitted core radius $r_{\mathrm{c}}$ of the DM halo induced by crossings of GCs with (in purple) and without (in green) a DM minihalo as a function of time. $r_{\mathrm{c}}\neq0$ ($r_{\mathrm{c}}=0$) means that there is a (no) cusp-to-core transition for the dwarf DM halo. This figure is adapted from \protect\cite{2021MNRAS.504.3509O,2020MNRAS.492.3169B}.}
\label{c1f99}
\end{figure}

Along with mechanisms that flatten the central DM density, there are mechanisms that can rebuild it. Even if feedback processes can generate cores in DM halos, simulations of dwarf galaxies have shown that a DM cusp could regenerate in the center of the halo \citep{2015MNRAS.449L..90L,2015MNRAS.454.2092O}. It was claimed that the infall of substructures such as minor mergers with cuspy halos is responsible for this cusp regrowth \citep{2015MNRAS.449L..90L,2015MNRAS.454.2092O,2003MNRAS.341..326D,2015MNRAS.454.2981C,2021MNRAS.504.3509O}. Figure~\ref{c1f99} illustrates the cusp regeneration of DM halo at late times due to the merger with a cuspy satellite. In the same way, it was shown that the passages of DM minihalos of GCs could significantly perturb the DM distribution in the Fornax halo centre \citep{2020MNRAS.492.3169B}. Indeed, between crossings, the halo can reform the cuspy halo owing to the new orbits of DM particles initially at the Fornax centre as they gained energy from the GCs. In the right panel of Figure~\ref{c1f99}, we observe reverse transitions of the Fornax DM halo. More precisely, there are forward and reverse transitions from the cusp to the core \citep{2020MNRAS.492.3169B}. We argue that DM minihalos, which are still orbiting in the host dwarf, induce potential fluctuations and then displace the DM potential centre. This potential shift is responsible for the cusp regeneration as these subhalos are much denser. As shown before, tidal interaction with a host galaxy can also contribute to core-cusp transformations. \cite{2010MNRAS.406.1290P} stressed that it is questionable whether DM cores in classical dwarfs could subsequently survive to the present day without being tidally disrupted by the MW. As such, the cuspy profile seems to be more common at recent epochs as predicted by \cite{2019MNRAS.484.1401R}. However it is unclear on which timescales this process is more likely to occur as it depends on the merger history and on the environment. As there is a cusp regrowth problem within CDM, this leaves open the question of maybe cores are only transient states. Therefore, we should expect to observe a diversity of DM profiles at a given mass. \cite{2019MNRAS.484.1401R,2015ApJ...810...22H} found that our local dwarf galaxies can be separated into two distinct classes, those with cold DM cusps and DM cores (see also Figure~\ref{c1f2}). This transient phenomenon could explain this diversity in the dwarf regime.

\subsubsection{The diversity problem}

\begin{figure}[!t]
\centering
\includegraphics[width=0.75\textwidth]{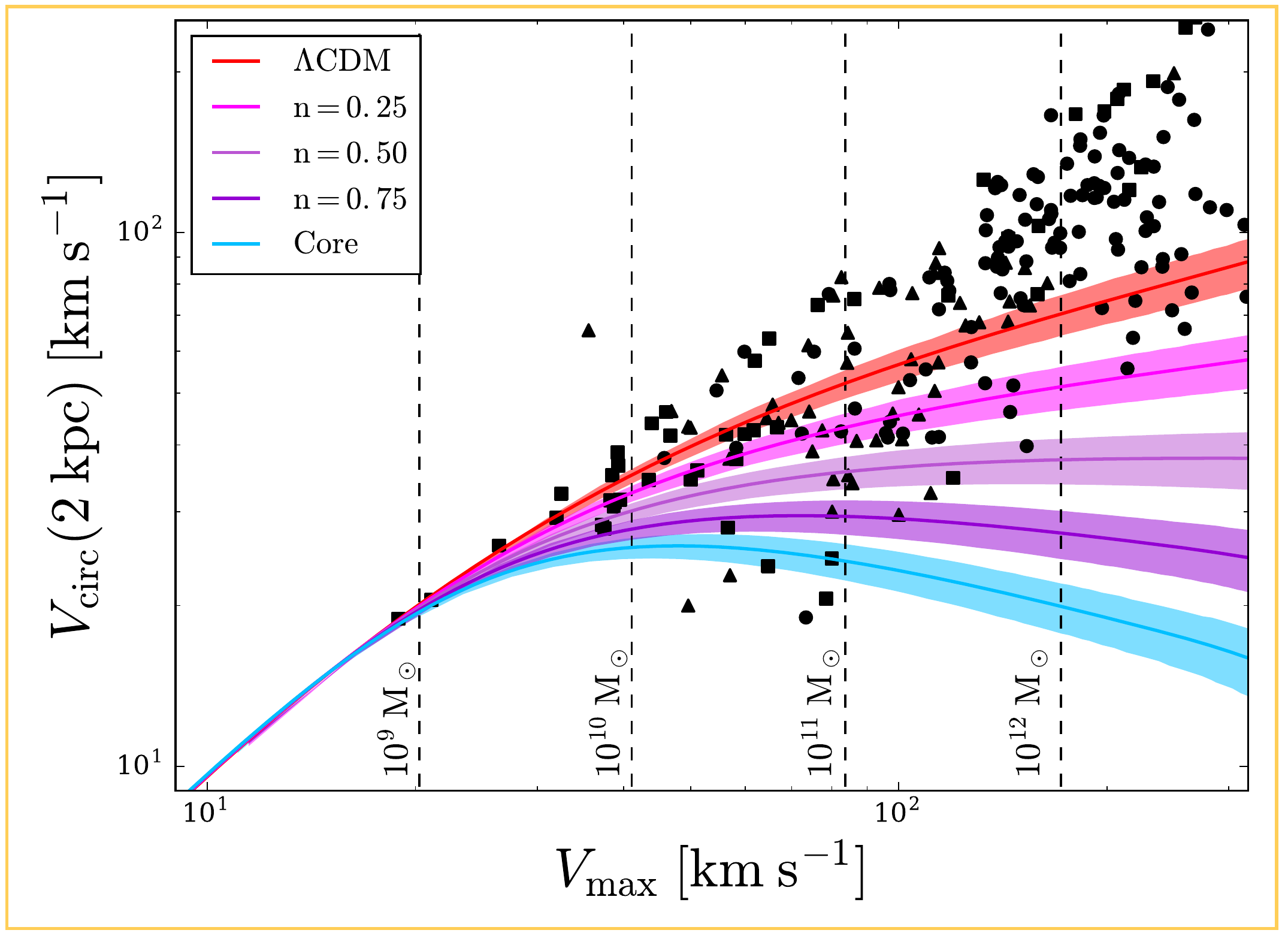}
\caption{{\it\bf Diversity problem:} Circular velocity at $r=2$ kpc versus the maximum circular velocity, $V_{\mathrm{max}}$ for observed galaxies. The lines trace the mean $V_{\mathrm{circ}}$(2 kpc) as a function of $V_{\mathrm{max}}$ described by $\Lambda$CDM (red), coreNFW model (see Equation~\eqref{eqcn}) for $n=0.25$ (magenta), $n=0.5$ (purple), $n=0.75$ (violet) and $n=1$ (blue), where the width of the bands correspond to the 1$\sigma$ scatter in DM halo concentrations (see Equation~\eqref{eqn2}). Observed galaxies with their observation type such as HI (black square), H$\alpha$ (black circle), and HI+H$\alpha$ (black triangle) were taken from the compilation by \protect{\cite{2015MNRAS.452.3650O}}. Galaxies below the red band are those with less mass within 2 kpc than expected from the predicted $\Lambda$CDM model.}
\label{c1f3}
\end{figure}

A key observable related to the inner mass distribution of galaxies is their rotation curve. The circular velocity curves of simulated galaxies vary systematically as a function of their maximum circular velocity $V_{\mathrm{max}}$ with a marginal uncertainty according to the CDM model. On the other hand, observed galaxies show a large diversity of rotation curve shapes, even at fixed maximum rotation velocity, especially for dwarf galaxies. This is at odds with the expectation for CDM halos, where $V_{\mathrm{max}}$ fully determines $V_{\mathrm{circ}}$(2 kpc) and it has been termed the diversity problem \citep{2015MNRAS.452.3650O}. The origin of this diversity is still not well understood. 

\medskip

Figure~\ref{c1f3} shows the circular velocity $V_{\mathrm{circ}}$(2 kpc) versus $V_{\mathrm{max}}$ for observed galaxies. We used the coreNFW model in order to characterize the inner DM density from these observed circular velocities. The coreNFW profile is a fitting function, which captures the cusp-core transformation \citep{2016MNRAS.459.2573R}. For this model, the cumulative mass profile is given by:
\begin{equation}
    M_{\mathrm{cNFW}}(<r)=M_{\mathrm{NFW}}(<r)f^n,
\end{equation}
where $M_{\mathrm{NFW}}$ is the NFW mass profile and $f^n$ generates a shallower density profile below a core radius $r_{\mathrm{c}}$:
\begin{equation}
    f^n=\left[\mathrm{tanh}\left(\frac{r}{r_\mathrm{c}}\right)\right]^n,
\end{equation}
where the parameter $0< n \leq 1$ controls how shallow the core becomes and corresponds to the transition region between cusp and core. Indeed, $n=0$ ($n=1$) corresponds to a fully cuspy (core) halo. The density profile of the coreNFW model is given by:
\begin{equation}
    \rho_{\mathrm{cNFW}}=f^n\rho_{\mathrm{NFW}} + \frac{n f^{n-1}(1-f^2)}{4\pi r^2r_{\mathrm{c}}}M_{\mathrm{NFW}}.
\label{eqcn}
\end{equation}
Given the halo mass and redshift, both halo concentrations $c_{200}$ can be estimated from cosmological $N$-body simulations. Indeed, the mass and concentration of halos at redshift $z=0$ in $\Lambda$CDM are correlated:
\begin{equation}
\mathrm{log_{10}}(c_{\mathrm{200}}) = 0.905 - 0.101 \log_{10}(M_{200}h - 12), 
\label{eqn2}
\end{equation}
with a scatter $\Delta \mathrm{log_{10}}(c_{\mathrm{200}})=0.1$, where $h$ is the Hubble parameter \citep{2014MNRAS.441.3359D}. In Figure~\ref{c1f3}, the lines trace the mean of the circular velocity at $r=2$ kpc as a function of $V_{\mathrm{max}}$. Galaxies below the red band are those with less mass within 2 kpc than expected from the predicted $\Lambda$CDM model. This is evidence for the presence of cores in such galaxies (see Figure~\ref{c1f3}). However, galaxies at large masses tend to have a higher circular velocity at $r=2$ kpc than expected from $\Lambda$CDM. This is explained by the non-negligible contribution of the baryons to the inner rotation curve in massive galaxies. We also note that the scatter in the circular velocity at 2 kpc is reduced for galaxies below the red band as well as the mass increase (see Figure~\ref{c1f3}). 

\medskip

Explaining this observed diversity demands a mechanism that creates cores of various sizes in only some galaxies, but not in others, over a wide range of $V_{\mathrm{max}}$. Nevertheless, these galaxies, formed in similar halos, have approximately the same baryonic mass, and similar morphologies. Some diversity induced by differences in the distribution of the baryonic component was expected, but clearly the observed diversity is much greater than in simulations \citep{2010Natur.463..203G,2011AJ....141..193O,2012MNRAS.424.1275B,2013MNRAS.429.3068T,2014ApJ...789L..17M}. Further, we would expect that the DM is most affected in systems where baryons play a more important role such as high-surface brightness galaxies, whereas observations seem to suggest the opposite trend \citep{2019MNRAS.488.2387B}. The observed diversity could be explained from the cusp regeneration phenomenon or from a different DM nature. Indeed, this behaviour of observed rotation curves is predicted by MOND theory \citep{2020Galax...8...35M}.

\subsection{Inherent cores}

The presence of the core appears to persist for dwarf galaxies that are DM dominated and baryon deficient. Thus, it is still unclear which dynamical process in a CDM environment can solve this puzzle. Another possibility is that the DM is more complex and hotter than simple CDM. A wide range of alternative DM models has been proposed over the last decades. Mostly three main classes of alternative DM models have been simulated: warm dark matter (WDM) \citep{2000ApJ...542..622C,2001ApJ...556...93B,2012MNRAS.424..684S,2012MNRAS.424.1105M,2013MNRAS.430.2346S,2014MNRAS.439..300L}, self-interacting dark matter (SIDM) ( \cite{2000ApJ...543..514K,2002ApJ...564...60M,2013MNRAS.431L..20Z,2015MNRAS.453...29E,2000ApJ...534L.143B,2000PhRvL..84.3760S} and for reviews see \cite{2017PhRvD..95d3541H,2021JCAP...01..011H}), and fuzzy dark matter (FDM) that fundamentally change the gravitational law \citep{2019PhRvL.123n1301M,2019MNRAS.482.3227N,2014MNRAS.437.2652M,2017PhRvD..95d3541H,2018PDU....22...80C,2000PhRvL..85.1158H,2014NatPh..10..496S,2000NewA....5..103G,2000ApJ...534L.127P}. Many of these alternative theories have been invoked to address $\Lambda$CDM small-scale problems and more particularly, the cusp-core problem. FDM and SIDM, which are the two most recent alternatives theories, are reviewed in this section. Besides, primordial black holes have recently been proposed to explain the cusp-core problem as from a dynamical perspective they behave like any other CDM candidate \citep{2020MNRAS.492.5218B}.

\subsubsection{Fuzzy dark matter}

\begin{figure}[!t]
\centering
\includegraphics[width=0.75\textwidth]{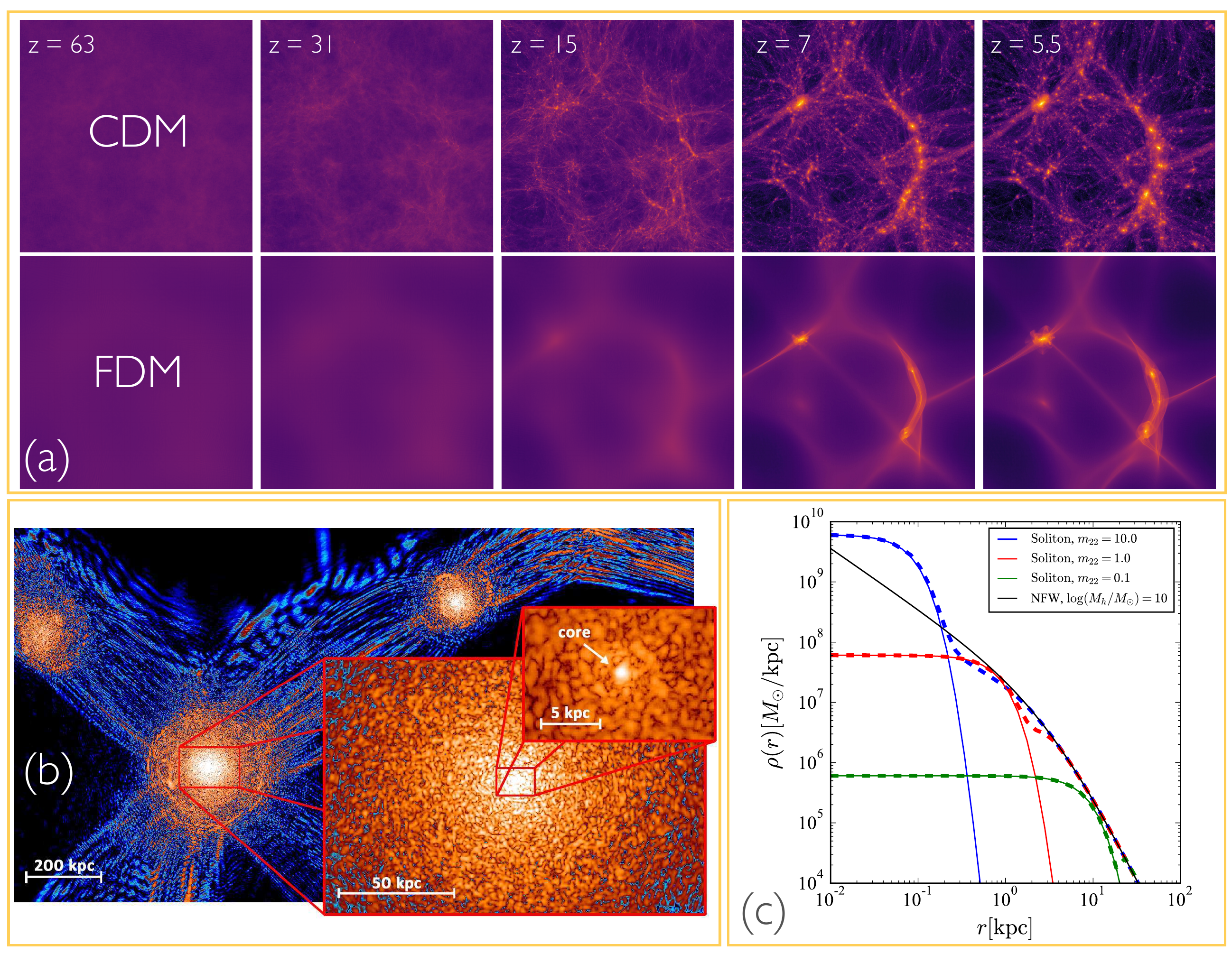}
\caption{{\it\bf FDM properties:} (a) Snapshots of the DM projected densities along the line of sight at $z=$ 63, 31, 15, 7, and 5.5 under the CDM (upper panel) and FDM (lower panel) cosmologies. The two cosmological simulations have led to the formation of three $\sim10^9-10^{10}$ M$_{\odot}$ halos. Snapshots highlight that FDM halos are connected via filaments, while CDM has filaments fragmented into subhalos. (b) Slice of density field of FDM simulation at different scales at $z=0.1$. We can distinguish the cores with a size of $\sim 0.3-1.6$ kpc in each halo. These DM cores grow as particles are accreted and surrounded by virialized halos. (c) DM profiles of a $10^{10}$ M$_{\odot}$ halo at $z=0$ for different values of $m_{22}$. Thin solid lines show the FDM core profiles for different axion masses. The thin black line shows the NFW profile of a $10^{10}$ M$_{\odot}$ halo at $z=0$. The thick dashed lines show the full halo profile that is a combination of the FDM profile transitioning to an NFW profile around $r=3r_{\mathrm{c}}$. This figure is adapted from \protect\cite{2014NatPh..10..496S,2020ApJ...893...21S,2019PhRvL.123n1301M}.}
\label{c2f1}
\end{figure}

As there is a current lack of evidence for any CDM particle such as weakly interacting massive particles, DM as an ultralight scalar field with no self-interaction in the non-relativistic limit was introduced under the name of Fuzzy Dark Matter (FDM) \citep{2000NewA....5..103G,2000PhRvL..85.1158H}. This scalar field is assumed to be made of very light particles with a mass of $\sim10^{-22}-10^{-19}$ eV. One of the candidates for this alternative DM theory is the axion-like particles predicted by string theories \citep{2016PhR...643....1M}. Such a scalar field is then well-described in the non-relativistic limit by the coupled Schrodinger and Poisson equations \citep{1993ApJ...416L..71W}:
\begin{equation}
    i\hbar \frac{\partial}{\partial t}\phi=-\frac{\hbar^2}{2m}\nabla^2\phi + mU\psi,
\end{equation}
\begin{equation}
    \nabla^2U=4 \pi G \rho_{\mathrm{m}},
\end{equation}
where $m$ is the mass of FDM particles. The mass density defines as $\rho_{\mathrm{m}}=|\phi|^2$ and $U$ is the gravitational potential. Such ultra-light DM particles have a characteristic wavelength called the de Broglie wavelength: 
\begin{equation}
    \lambda= 1.19 \left(\frac{10^{-22} \mathrm{eV}}{m}\right)\left(\frac{100 \mathrm{km.s}^{-1}}{v}\right) \mathrm{kpc},
    \label{eqz1}
\end{equation}
where $v$ is the characteristic velocity. Equation~\eqref{eqz1} shows that the wavelength of a few kpc is of astrophysical size. Indeed, the small masses of ultra-light DM particles are associated with a very large de Broglie wavelength where their quantum properties play an important role \citep{2000PhRvL..85.1158H,2009ApJ...697..850W,2007JCAP...06..025B}. Thus, the de Broglie wavelength is of the order of the scales at which the cusp-core problem appears. 

\medskip
Axion-like particles are interesting DM candidates because they predict new structural and dynamical phenomena on scales of galaxies. When the de Broglie wavelength $\lambda$ is on the order of or larger than the inter-particle distance $d_{i}$, quantum effects will dominate. In fact, DM particles have huge occupancy numbers at these small scales. In the non-interacting Bose gas theory, the macroscopic occupation of the ground state is seen as condensation and this phenomenon is called Bose-Einstein condensation. In FDM, the particles form a Bose-Einstein condensate on galactic scales \citep{2000PhRvL..85.1158H}. Figure~\ref{c2f1} depicts that it results in a DM core at the halo's central region as the particles of the system are in the ground state described by a single wave function \citep{2007JCAP...06..025B,2014NatPh..10..496S}. Cosmological simulations of light DM found that the density profile of the innermost central region of the halos at redshift $z=0$ follows \citep{2014NatPh..10..496S}:
\begin{equation}
    \rho(r)=\frac{\rho_0}{\left(1+0.091(r/r_{\mathrm{c}})^2\right)^8}10^9 M_{\odot} \mathrm{kpc}^{-3}, 
\end{equation}
with 
\begin{equation}
    \rho_0=0.019 m_{22}^{-2} r_{\mathrm{c}}^{-4} 10^9 M_{\odot} \mathrm{kpc}^{-3},
\end{equation}
where $m_{22}=m/10^{-22}$ eV is the DM particle mass and $r_{\mathrm{c}}$ is the radius at which the density drops to one-half its peak value for a halo at $z=0$. The central mass density of the core is given by \cite{2019MNRAS.483..289R}:
\begin{equation}
    M_{\mathrm{c}}=\frac{M_{\mathrm{h}}^{1/3}}{4}\left(4.4\times10^7 m_{22}^{-3/2} \right)^{2/3},
    \label{mc}
\end{equation}
and 
\begin{equation}
    r_{\mathrm{c}}=\frac{1.6}{m_{22}}\left(\frac{M_{\mathrm{h}}}{10^9 M_{\odot}}\right)^{-1/3} \mathrm{kpc},
\end{equation}
where $M_{\mathrm{h}}$ is the halo mass. The heating mechanism is due to quantum fluctuations arising from the uncertainty principle. Indeed, the quantum pressure stabilizes the gravitational collapse and prevents the formation of cusp by suppressing the small-scale structures \citep{2009ApJ...697..850W,2000PhRvL..85.1158H,2010JCAP...01..007L}. The condensate is a stable region where no clustering takes place (see Figure~\ref{c2f1}). These kpc cores offer one possible solution to the cusp-core problem. 

\medskip

However, when $\lambda \ll d_{\mathrm{i}}$, DM particles can be considered to be a classical system. Indeed, at large scales, condensation is broken and the system behaves as a system of individual massive particles \citep{2019PhRvL.123n1301M}. Figure~\ref{c2f1} shows that the outer region of FDM halo behaves like CDM which is well approximated by the NFW \citep{2014NatPh..10..496S}. Thus, the full density profile of halos can be written as:
\begin{equation}
    \rho(r) = \Theta(r_t - r)\rho_c + \Theta(r_t - r)\rho_{NFW},
\end{equation}
where the $\Theta$ is a step function and $r_{\mathrm{t}}$ is the transition radius, which marks the transition between the core profile and NFW profile. This specific scale is proportional to the core size as $r_{\mathrm{t}} = \alpha r_{\mathrm{c}}$ where $\alpha\sim 2-4$ \citep{2019MNRAS.483..289R}.

\medskip

\begin{figure}[!t]
\centering
\includegraphics[width=0.75\textwidth]{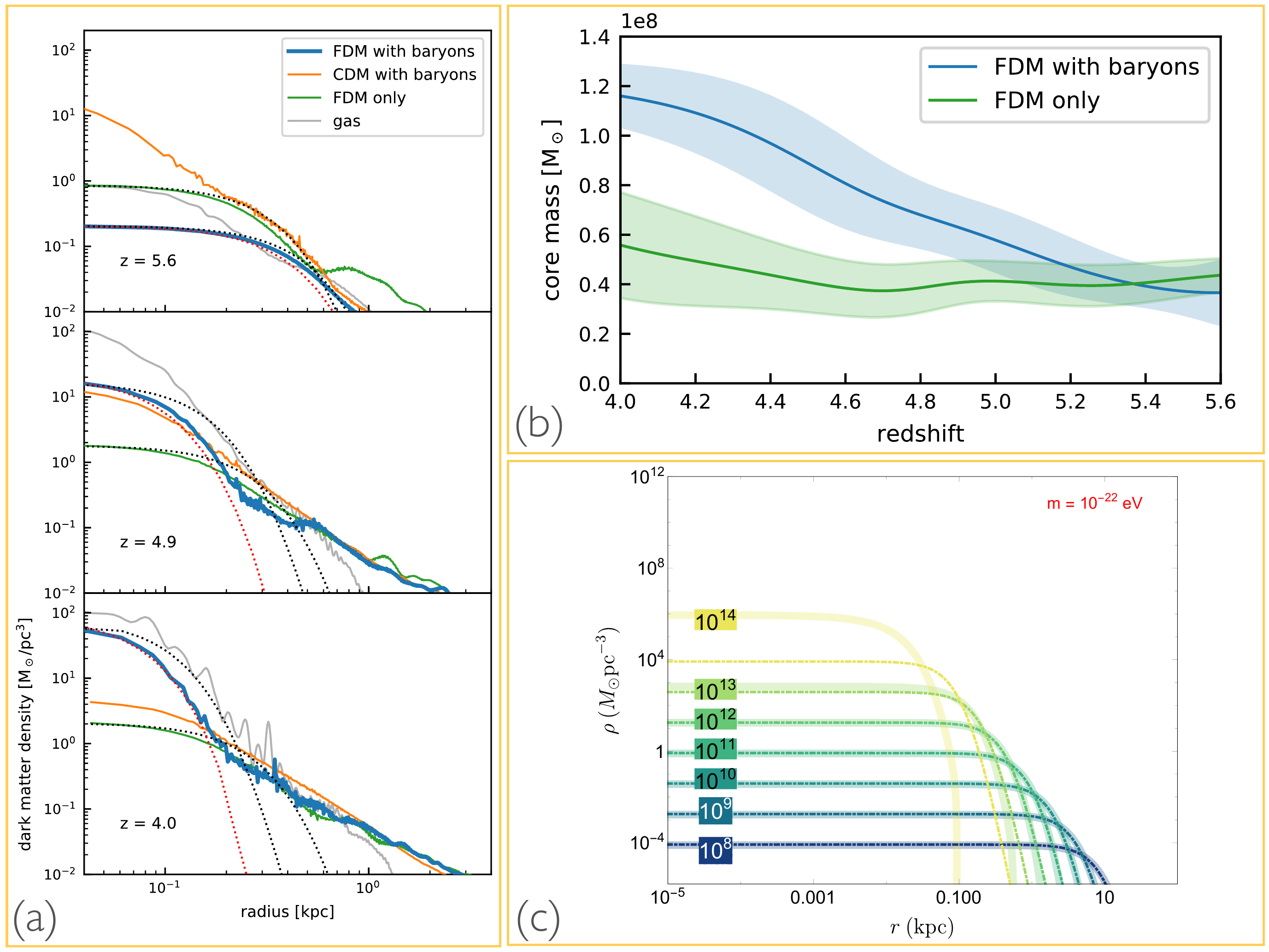}
\caption{{\it\bf Adding baryons and black holes:} (a) DM density profiles of a halo with a final virial mass of $10^{10}$ M$_{\odot}$ at three different redshifts assuming FDM only, FDM with baryons and CDM with baryons. The FDM halo with baryons has a lower density at the centre than the FDM-only halo because the baryon pressure delays its collapse. (b) Evolution of the core mass over redshift for a FDM halo with (blue curve) and without baryons (green curve). The shaded regions represent the corresponding standard deviations. In the presence of baryons, the cores grow by more than a factor of two. (c) Density profiles of FDM halos with masses from $10^8$ M$_{\odot}$ up to $10^{14}$ M$_{\odot}$ assuming a FDM particle mass of $m=10^{-22}$ eV. The dot-dashed (thicker) lines correspond to DM halos without (with) a central black hole. It can be seen that the black hole increases the central density only for $M_{\mathrm{h}} \ge 10^{13}$ M$_{\odot}$. This figure is adapted from \cite{2020PhRvD.101h3518V,2020MNRAS.492.5721D}.}
\label{c2f2}
\end{figure}

\begin{figure}[!t]
\centering
\includegraphics[width=0.75\textwidth]{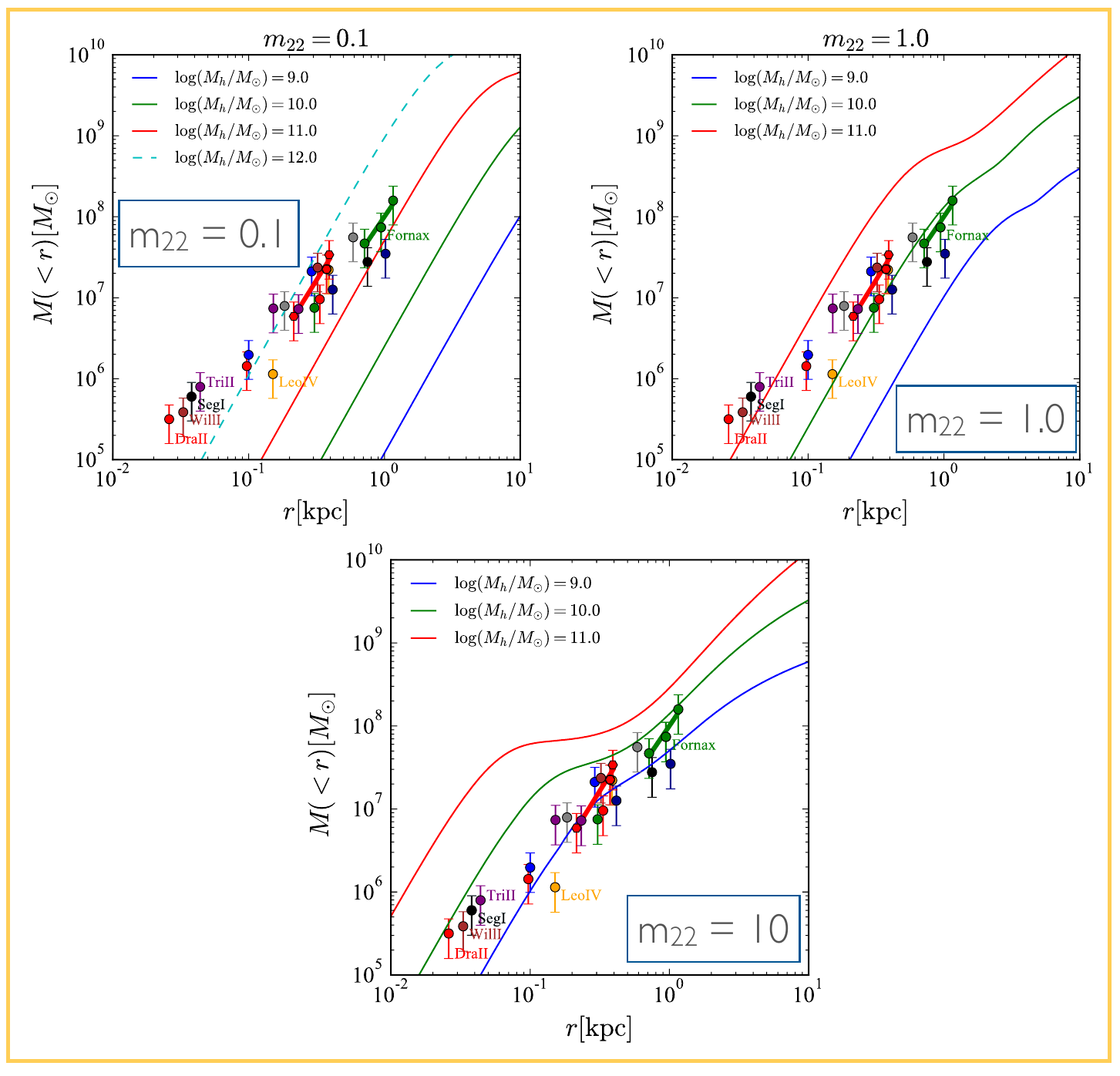}
\caption{{\it\bf FDM inconsistencies:} Mass profiles of FDM halos with masses from $10^9$ M$_{\odot}$ up to $10^{11}$ M$_{\odot}$. Left, middle, and right panels represent the mass profiles corresponding to $m_{22}=$0.1, 1, and 10, respectively. The individual data points for dwarf galaxies are collected from \cite{2010MNRAS.406.1220W,2016ApJ...818...40M}, and the slopes of Fornax and Sculptor (green and red lines) are from \cite{2011ApJ...742...20W}. The profiles show a core region parametrized by Equation~\eqref{mc}\label{section6} which then follow an NFW profile at $r=3r_{\mathrm{c}}$. These plots highlight the tensions concerning the FDM particle mass. This figure is adapted from \cite{2020ApJ...893...21S}.}
\label{c2f3}
\end{figure}

FDM was introduced by the motivation to solve the core-cusp problem in DM halos of galaxies. As halo cores form naturally in FDM theory, this scenario is appealing in principle. However, some specific observations are necessary to verify this type of DM. The quantum nature of DM particles gives rise to specific density profiles and potential fluctuations that may affect delicate structures such as tidal streams and disks \citep{2019JPhCS1253a2007E}. 

\medskip

As illustrated above, any DM model, which sets a universal core profile cannot fit observations. As such, baryonic physics must also play a significant role in shaping the DM profiles. Figure~\ref{c2f2} illustrates DM density profiles of a halo with a final virial mass of $10^{10}$ M$_{\odot}$ at three different redshifts assuming FDM, FDM with baryons and CDM with baryons. At the earliest redshift $z=$5.6, the CDM halo exhibits the highest central DM density with a cuspy profile, while the FDM halos show core profiles. The FDM halo with baryons has a lower density at the centre than the FDM-only halo because the baryon pressure delays its collapse. In contrast, at $z=$4, the FDM central density is more than one order of magnitude higher with baryons than without, exceeding the central DM density of the CDM halo. Indeed, in the presence of baryons, the cores grow by more than a factor of two. However, the core mass does not evolve over time if baryons are absent (see Figure ~\ref{c2f2}). As DM cores become more massive and compact in the presence of baryons, observed rotation curves are likely harder to reconcile with FDM \citep{2020PhRvD.101h3518V}.

\medskip

Moreover, we expect that DM distribution of centrally baryon-dominated galaxies, especially those containing supermassive black holes, are more strongly affected \citep{2019JCAP...07..045B,2019MNRAS.488.4497D,2019PhRvL.123b1102D,2020MNRAS.492.5721D}. Figure~\ref{c2f2} also shows the density profiles of FDM halos with masses from $10^8$ M$_{\odot}$ up to $10^{14}$ M$_{\odot}$ assuming a FDM particle mass of $10^{-22}$ eV. It can be seen that the black hole increases the central density only for $M_{\mathrm{h}}\ge10^{13}$ M$_{\odot}$. This latter effect depends also on the FDM particle mass. Thus, black holes are most effective at modifying the DM distribution for higher halo masses, and larger FDM particle masses. By numerically solving the Schrodinger-Poisson equations, it was shown that black holes decrease the core radius by increasing the central density of DM halos (see Figure~\ref{c2f2}).

\medskip

Moreover, it not clear if FDM halos can be in line with known galaxy scaling relation \citep{2019JCAP...07..045B,2018PhRvD..98b3513D,2020ApJ...893...21S,2019MNRAS.483..289R}. Figure~\ref{c2f3} depicts the mass profiles of FDM halos with masses from $10^9$ M$_{\odot}$ up to $10^{11}$ M$_{\odot}$. For $m_{22}=$0.1, the predicted halo mass of the dwarf galaxies is too high given their dynamical state in the galaxy, and higher $m_{22}$ does not agree with the inferred slopes of Sculptor and Fornax. Low mass axions ($m_{22}=$0.1) can explain the observed mass profile slopes in Sculptor and Fornax \citep{2017MNRAS.472.1346G,2015MNRAS.451.2479M,2014PhRvL.113z1302S}. However, at such low masses, the predicted halo masses of the ultra-faint dwarf galaxies such as Segue I are ruled out by dynamical friction arguments. In contrast, high mass axions ($m_{22}=$10) can explain the halo masses of the ultra-faint dwarf galaxies such as Draco II, Triangulum II, and Segue I. For this axion mass, the predicted mass profiles do not agree with the observed slope of Fornax and Sculptor. The latter highlights the tensions concerning the FDM particle mass (see Figure~\ref{c2f3}). Indeed, stellar velocity measurements around the central BH of the MW constrained the FDM particle mass to be $m<10^{-19}$ eV \citep{2019JCAP...07..045B}. However, the central motion of bulge stars in the MW favors a mass of $10^{-22}$ eV \citep{2020PDU....2800503D}. A similar value was found by applying a Jeans analysis for MW dwarfs \citep{2017MNRAS.468.1338C}. 
\medskip

Besides, it was pointed out that observational data indicate a positive scaling between the core radius $r_{\rm c}$ and the halo mass $M_{\rm h}$ \citep{2020arXiv200503520D,2000ApJ...537L...9S}. In other words, we expect to have larger cores in massive galaxies. However, the FDM theory seems to predict the opposite behaviour. Indeed, the core radius is a decreasing function of the halo mass in FDM universe, expressed
as $r_{\rm c} \propto M^{\alpha}$ with $\alpha=1/3-5/9$ \citep{2020ApJ...904..161B,2020arXiv200704119M,2021arXiv211011882J,2017MNRAS.471.4559M,2014NatPh..10..496S,2016PhRvD..94d3513S,2021MNRAS.501.1539N}. Thus, it seems very difficult for FDM to reproduce the observed relationship between core radius and halo mass in galaxies. To sum up, FDM model provide a natural framework for the formation of DM cores but its predictions are in conflict with observations of galaxies.

\subsubsection{Self-interacting dark matter}

\begin{figure}[!t]
\centering
\includegraphics[width=0.75\textwidth]{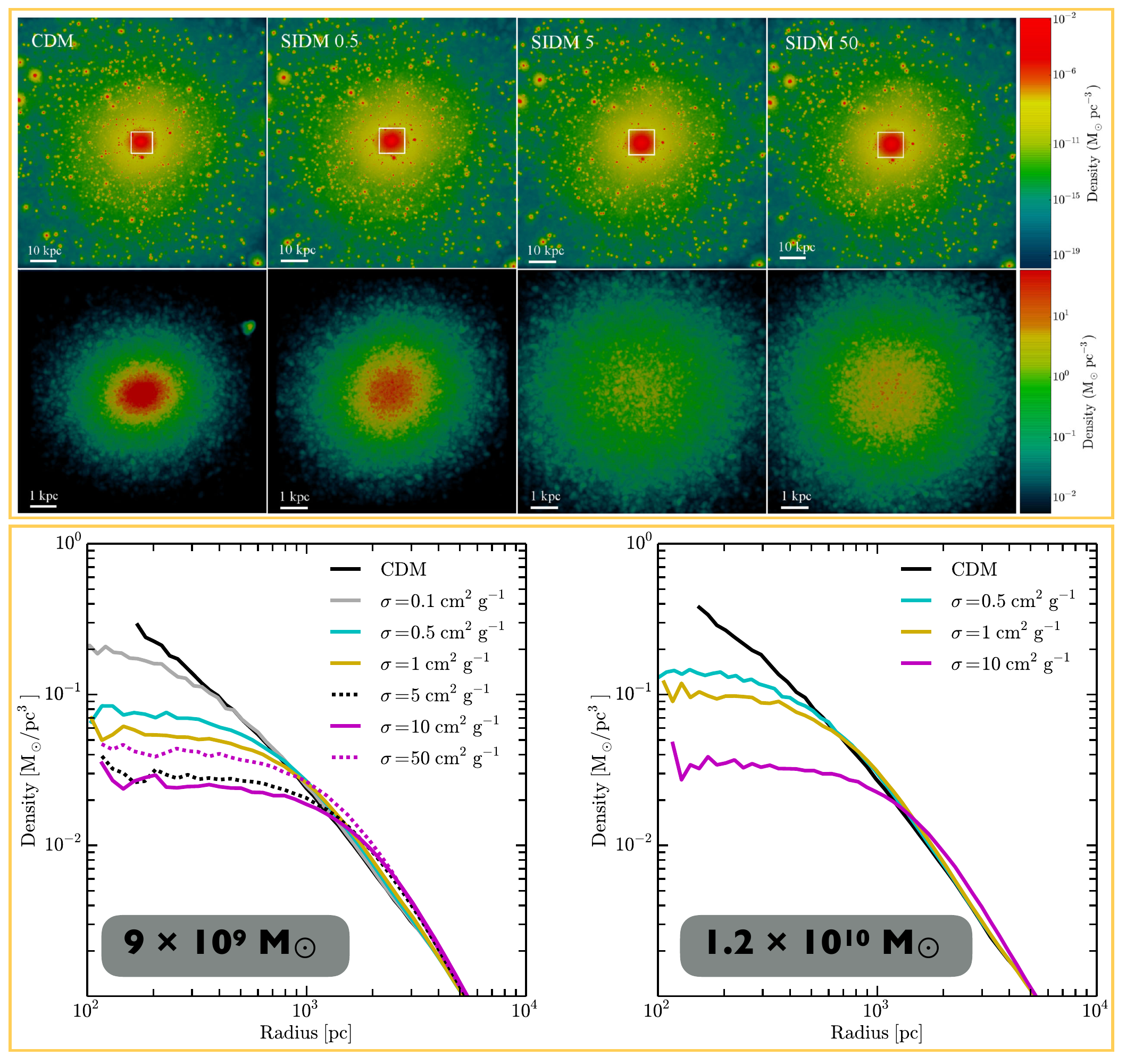}
\caption{{\it\bf SIDM halo properties:} {\it Upper panel}: DM density maps of CDM (left) and SIDM halo with a mass of $9\times10^9$ M$_{\odot}$ and $\sigma/m$ increasing from left to right in 100 and 10 kpc boxes. SIDM halos have the same structure as CDM halos at large scales. At sub-galactic scales, the SIDM halos are less dense than in CDM model due to the formation of cores. {\it Lower panel}: DM density profiles of $9\times10^9$ (left) and $1.2\times10^{10}$ M$_{\odot}$ (right) halos in CDM and SIDM models. SIDM runs have $\sigma/m$ between 0.1 and 50 cm$^2$ g$^{-1}$. For $\sigma/m \geq 0.5$ cm$^2$ g$^{-1}$, the self-interactions between DM particles produce central cores with a size depending on $\sigma/m$. This figure is adapted from \cite{2015MNRAS.453...29E}.}
\label{c3f1}
\end{figure}

\begin{figure}[!b]
\centering
\includegraphics[width=0.75\textwidth]{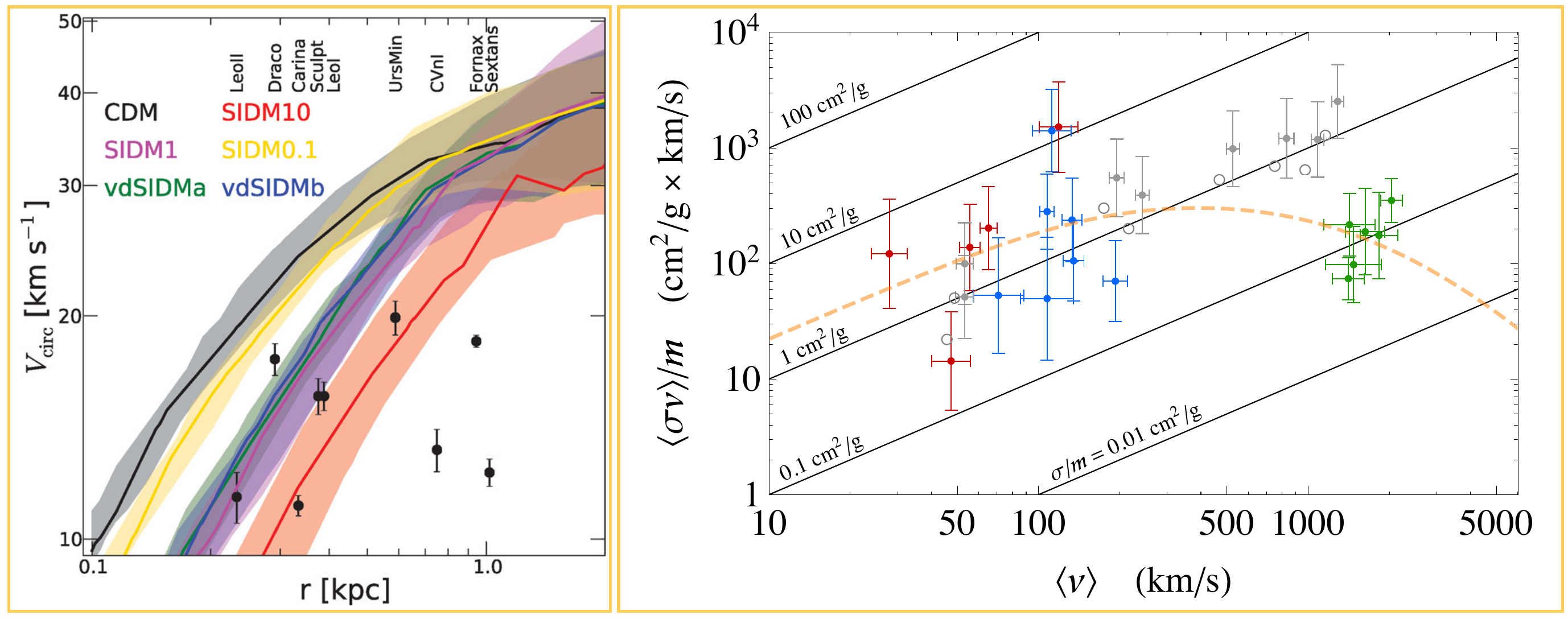}
\caption{{\it\bf Constraints from observations:} Left panel: Circular velocity profiles encompassing a distribution of 15 subhalos for CDM and SIDM models with a constant cross-section between 0.1 and 10 cm$^2$ g$^{-1}$. Black points with error bars correspond to the circular velocity within the half-light radii for nine MW dSphs \citep{2010MNRAS.406.1220W,2009ApJ...704.1274W}. While the most massive CDM subhalos are inconsistent with the kinematics of the MW dSphs, the SIDM model with $\sigma/m >1$ cm$^2$ g$^{-1}$ can alleviate this problem. Right panel: Velocity-weighted cross-section per unit mass as a function of the mean collision velocity for dwarf galaxies (red), Low surface brightness (LSB) galaxies (blue) and galaxy clusters (green). For comparison, SIDM $N$-body simulations with $\sigma/m = 5-10$ cm$^2$ g$^{-1}$ are represented by grey points. Diagonal lines show the corresponding cross-section $\sigma/m$. As $\sigma/m$ is not supposed to be constant in velocity, it is more convenient to invoke $<\sigma v>/m$ rather than $\sigma/m$. The dashed curve represents the best-fit for a velocity-dependent cross-section. This figure is adapted from \cite{2016PhRvL.116d1302K}.}
\label{c3f2}
\end{figure}

\begin{figure}[!t]
\centering
\includegraphics[width=0.75\textwidth]{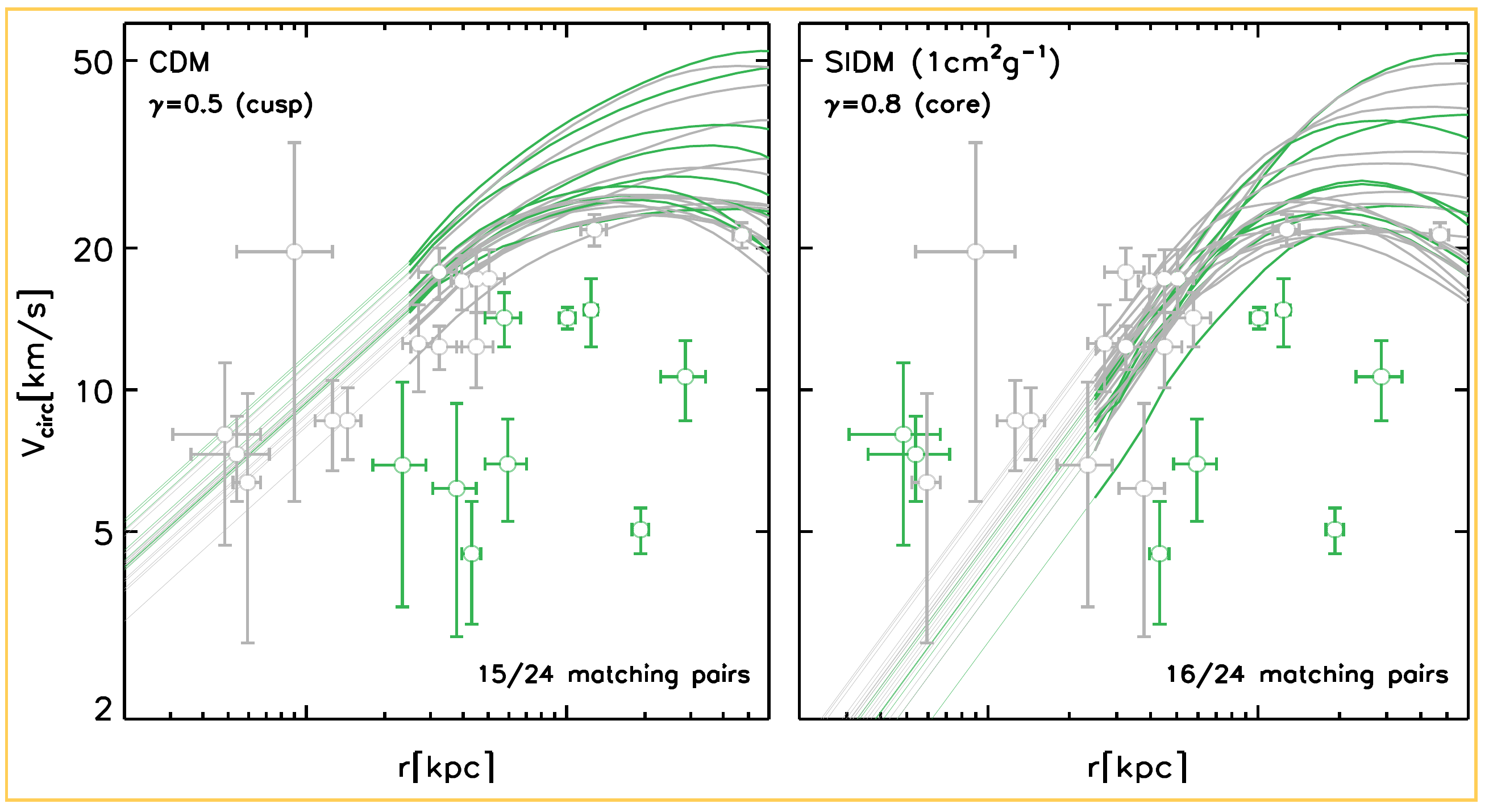}
\caption{{\it\bf SIDM versus CDM:} Circular velocity profiles $V_{\mathrm{circ}}$ of CDM and SIDM ($\sigma/m = 1$ cm$^2$ g$^{-1}$) subhalos within 300 kpc from the centre of the simulated MW-like galaxies. Open symbols with error bars correspond to circular velocities at the half-light radius for 24 MW satellites \citep{2018MNRAS.481.5073E,2019MNRAS.488.2743T}. Lines and symbols in gray (green) are consistent matches (mismatches) between simulated subhalos and data points. Both CDM and SIDM subhalos match only 15-16 MW satellites. This figure is adapted from \cite{2019PhRvD.100f3007Z}.}
\label{c3f3}
\end{figure}

\begin{figure}[!t]
\centering
\includegraphics[width=0.75\textwidth]{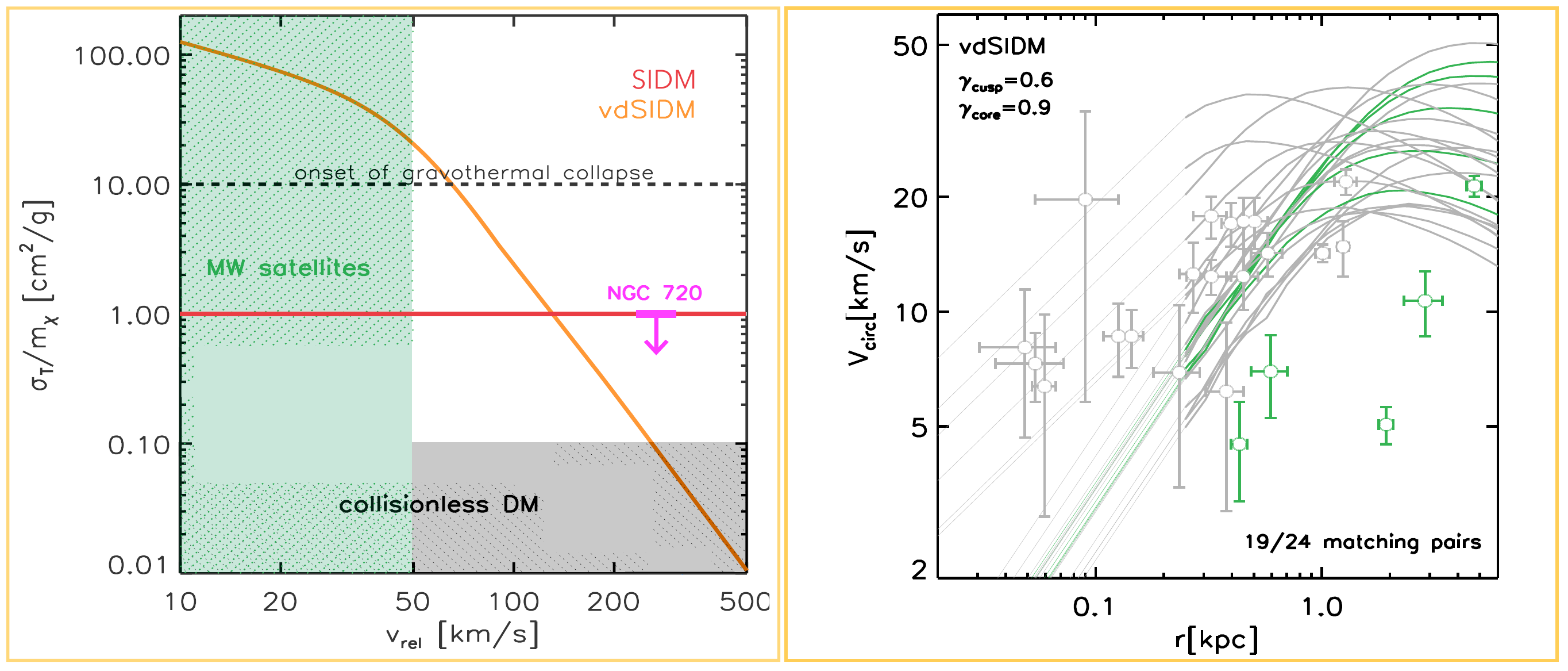}
\caption{{\it\bf Velocity-dependent cross-section:} Left panel: Cross-section as a function of the relative velocity. vdSIDM model consists of a SIDM with a strong velocity-dependent cross-section (orange line). The collisionless region is delimited by the black area $\sigma/m < 0.1$ cm$^2$ g$^{-1}$. For $\sigma/m > 10$ cm$^2$ g$^{-1}$, self-interactions between DM particles are frequent enough to result in core-collapse within a Hubble time in halos. The green area represents the relevant region for MW satellites. A constraint on the cross-section from the elliptical galaxy NGC720 is represented by a magenta arrow. Right panel: Circular velocity profiles $V_{\mathrm{circ}}$ of vdSIDM (orange line in the left panel) subhalos within 300 kpc from the centre of the simulated MW-like galaxies. Open symbols with error bars correspond to circular velocities at the half-light radius for 24 MW satellites \citep{2018MNRAS.481.5073E,2019MNRAS.488.2743T}. Lines and symbols in gray (green) are consistent matches (mismatches) between simulated subhalos and data points. As the vdSIDM model has cross-sections near and above the core-collapse limit, it produces a bimodal distribution composed of cusps and cores for MW-like subhalos. This figure is adapted from \cite{2019PhRvD.100f3007Z}.}
\label{c3f4}
\end{figure}

In the $\Lambda$CDM model, DM is assumed to be collisionless. Another promising alternative is, therefore, self-interacting dark matter (SIDM) \citep{1995ApJ...452..495D,1992ApJ...398...43C}, proposed to solve the small scales problems, and more specifically the cusp-core problem \citep{2000PhRvL..84.3760S}. In this scenario, it was initially assumed that DM interactions are isotropic elastic scatterings with an interaction cross-section that is independent of velocity. Since the mass of the DM particle is not known, self-interactions are commonly quantified in terms of the cross-section per unit particle mass, $\sigma/m$, which is an important cosmological value for SIDM theories. The total number of interactions, $\Gamma$, that occurs per unit time is given by
\begin{equation}
   \Gamma \sim 0.1\, \mathrm{Gyr^{-1}} \times \left(\frac{\rho_{\mathrm{dm}}}{0.1 M/\mathrm{pc^3}} \right) \left( \frac{\sigma/m}{1\, \mathrm{cm^2 g^{-1}}}\right)\left(\frac{v_{\mathrm{rel}}}{50\,\mathrm{km s^{-1}}}\right),
\label{eqs1}
\end{equation}
where $m$, $\sigma$ and $v_{\mathrm{rel}}$ are the DM particle mass, the cross-section, and the relative velocity, respectively. The upper panel of Figure~\ref{c3f1} compares the DM density distribution at large scales of CDM and SIDM halos. As the scattering rate $\Gamma$ is proportional to the DM density, SIDM halos have the same structure as CDM halos at large scales where the DM interactions are negligible. Indeed, on the scale of their virial radius ($r_{\mathrm{vir}}=55$ kpc), CDM and SIDM halos are nearly identical. Moreover, the collision rate is also negligible during the early Universe when DM structures form. Therefore, SIDM is consistent with observations of large-scale structures, predicted by $\Lambda$CDM \citep{2011ApJ...742...16T,2006Natur.440.1137S}. However, self-interactions perturb the inner density structure of DM halo at late times. The upper panel of Figure~\ref{c3f1} highlights that the SIDM halos at sub-galactic scales are less dense than in the CDM model due to the formation of cores.

\begin{figure}[!t]
\centering
\includegraphics[width=0.75\textwidth]{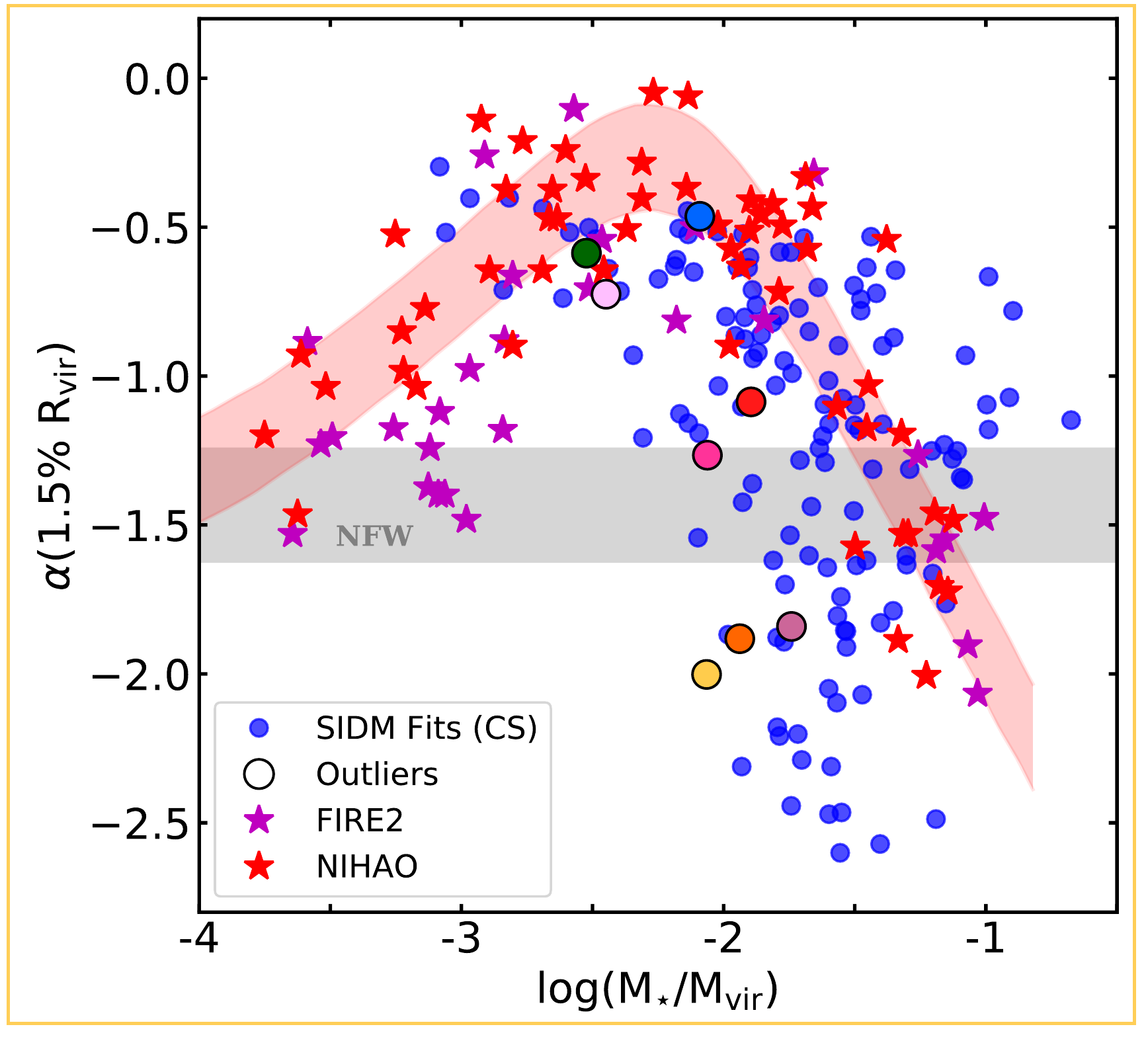}
\caption{{\it\bf Baryon impact on SIDM halos:} Inner DM density slope $\alpha$ at $r=$0.015$r_{\mathrm{vir}}$ as a function of $M_*/M_{\mathrm{vir}}$ at $z=0$ from SIDM fits \citep{2019PhRvX...9c1020R}, NIHAO \citep{2016MNRAS.456.3542T} and FIRE-2 \citep{2019MNRAS.490..962F,2018MNRAS.480..800H} hydrodynamical CDM simulations. The SIDM fits of the SPARC sample \citep{2016AJ....152..157L}, which contains 135 galaxies, including the impact of baryons on the halo profile and compatible with a unique cross-section of 3 cm$^2$ g$^{-1}$. The shaded grey band shows the expected range of DM profile slopes for the NFW profile as derived from CDM-only simulations by including concentration scatter. The slope $\alpha$ of SIDM fits spans a large range from -0.5 to -2.5, indicating that the SIDM model predicts both cored and cuspy halos. This figure is adapted from \cite{2019arXiv191100544K}.}
\label{c3f5}
\end{figure}

\medskip
A generic prediction for SIDM is that halos can form dense cores with size depending on the cross-section $\sigma/m$ \citep{2013MNRAS.431L..20Z,2014MNRAS.444.3684V,2015AAS...22540205F,2017MNRAS.465..569R,2017MNRAS.472.2945R,2013MNRAS.430...81R,2013MNRAS.430..105P,2015MNRAS.453...29E,2012MNRAS.423.3740V,2002ApJ...581..777C,2001ApJ...547..574D,2000ApJ...544L..87Y,2000ApJ...535L.103Y}, as shown in the lower panel of Figure~\ref{c3f1}. The redistribution of energy and momentum by DM particle collisions decreases the central density of DM halos, known as a cusp-to-core transition \citep{2000PhRvL..84.3760S,2000ApJ...535L.103Y,2013MNRAS.430...81R,2000ApJ...534L.143B,2003ApJ...586...12D}. In other words, this heating transfer alters the inner region of halos by turning cuspy profiles into cored profiles. Core formations occur only if $\sigma/m$ is sufficiently large to ensure that the relatively high probability of scattering over a time $T_{\mathrm{age}}$ is comparable to the age of the halo: $\Gamma \times T_{\mathrm{age}} \sim 1$. Figure~\ref{c3f1} illustrates that the self-interactions between DM particles produce central cores for $\sigma/m \geq 0.5$ cm$^2$ g$^{-1}$ in $9\times10^9-1.2\times10^{10}$ M$_{\odot}$ halos. Numerous simulations have then demonstrated that models with $\sigma/m \sim 0.5-10$ cm$^2/$ g$^{-1}$ produce DM cores in dwarf galaxies with sizes $\sim0.3-1.5$ kpc \citep{2012MNRAS.423.3740V,2013MNRAS.430..105P,2013MNRAS.430...81R,2013MNRAS.431L..20Z,2015AAS...22540205F,2015MNRAS.453...29E} that could alleviate the cusp-core problem. In fact, the discrepancy with observations of low surface brightness (LSB) galaxies having DM cores could be avoided in SIDM theory \citep{2000PhRvL..84.3760S}.

\medskip
The viability of DM self-interacting as a cusp-core transformation mechanism depends on whether or not this cosmological model is consistent with all observations. In fact, it remains to see if SIDM models are able to explain the observed cores from ultra-faint galaxies to galaxy clusters. SIDM model requires compromises on the cross-section, which needs to be small enough to be observationally allowed but sufficiently large to alleviate the relevant small-scale problem. The first constraint on the SIDM cross-section derives from galaxy clusters, which impose $\sigma/m < 0.02$ cm$^2/$g \citep{2002ApJ...564...60M}. Later, this constraint was revised and the inferred values of $<\sigma v>/m$ for all six clusters are consistent with a constant cross-section $\sigma/m = 0.1$ cm$^2$ g$^{-1}$ according to right panel of Figure~\ref{c3f2} \citep{2013MNRAS.430..105P,2018PhR...730....1T}. The left panel of Figure~\ref{c3f2} shows that SIDM model ($\sigma/m = 0.1$ cm$^2$ g$^{-1}$) allowed by cluster constraints would be very similar to the CDM predictions. While the most massive CDM subhalos are inconsistent with the kinematics of the MW dSphs, SIDM model can only alleviate this problem for $\sigma/m >1$ cm$^2$ g$^{-1}$. 

\medskip
If the self-scattering cross-section per unit mass is $\sim 1$ cm$^2$ g$^{-1}$, SIDM models can solve the cusp-core problem at the scale of dwarf galaxies \citep{2012MNRAS.423.3740V,2013MNRAS.430...81R,2013MNRAS.431L..20Z}. Figure~\ref{c3f3} depicts the circular velocity profiles $V_{\mathrm{circ}}$ of CDM and SIDM ($\sigma/m = 1$ cm$^2$ g$^{-1}$) subhalos. Both CDM and SIDM subhalos match only 15-16 MW satellites. Nevertheless, SIDM theory with constant cross-section ($\sigma/m = 1$ cm$^2$ g$^{-1}$) predicts DM subhalos with too low densities to match the observations of ultra-faint galaxies (see Figure~\ref{c3f3}). Thus, a constant cross-section of $\sigma/m = 1$ cm$^2$ g$^{-1}$ is likely to be inconsistent with the observed halo shapes of ultra-faint galaxies and several galaxy clusters. 

\medskip
Figure~\ref{c3f4} highlights the possible velocity dependence discernible in these data from dwarfs to clusters. As $\sigma/m$ varies within a wide range, SIDM models, which assume a constant scattering cross-section, need to be abandoned since those that could solve the cusp-core problem in dwarfs, seems to violate several astrophysical constraints. In order to alleviate the cusp-core problem and also match constraints at different scales, SIDM models need to have a velocity-dependent cross-section $\sigma(v)$ that decreases as the relative velocity of DM particles involves from dwarfs to clusters such as in Figure~\ref{c3f4} \citep{2012MNRAS.423.3740V,2011PhRvL.106q1302L,2010PhRvL.104o1301F}. For $\sigma/m > 10$ cm$^2$ g$^{-1}$, self-interactions between DM particles are frequent enough to entail a core-collapse, which is a well-known mechanism in globular clusters \citep{1968MNRAS.138..495L}, within a Hubble time in halos. Then, it results in the collapse of the core into a central cusp for SIDM halos \citep{2002ApJ...568..475B,2002ApJ...581..777C,2011MNRAS.415.1125K,2015ApJ...804..131P,2020PhRvD.101f3009N}. As the vdSIDM model has cross-sections near and above the core-collapse limit according to Figure~\ref{c3f4}, it produces a bimodal distribution composed of cusps and cores for MW-like subhalos. Indeed, the core collapse is responsible for this diversity, which is more consistent with cored brighter satellites and cuspy ultra-faint galaxies \citep{2019PhRvD.100f3007Z}. Thus, core collapses can be considered as a mechanism to create a diverse population of dwarf-size halos, some of which would be cuspy and others that would have cores in velocity-dependent SIDM models \citep{2019PhRvD.100f3007Z,2020PhRvD.101f3009N,2020PhRvL.124n1102S,2019JCAP...12..010K}.

\medskip

All previous works are based on SIDM simulations without taking into account baryonic physics. The inclusion of baryons into CDM simulations of dwarf galaxies has initially served to reduce the discrepancy between DM-only simulations and observations concerning the inner DM distribution. We have shown previously that baryonic feedback can reduce the central density of a cuspy DM halo. By including hydrodynamics in SIDM simulations, it was found that the DM inner region of dwarf galaxies with stellar masses $M_{*}<10^6$ are nearly identical to the SIDM-only simulations \citep{2014MNRAS.444.3684V,2015MNRAS.452.1468F,2017MNRAS.472.2945R}. Substantial DM cores are formed in both SIDM and SIDM+baryons simulations. It appears then that SIDM is more robust to feedback than CDM at dwarf scales \citep{2017MNRAS.472.2945R,2018ApJ...853..109E}. This suggests that the faintest dwarf spheroidals provide excellent laboratories constraining SIDM models. Indeed, they are ideal targets as SIDM and CDM produce cores and cusps in these galaxies, respectively. 

\medskip
For high baryon concentration, it leads to a dense inner halo with a smaller core in SIDM model \citep{2014PhRvL.113b1302K}. Moreover, baryons can cause SIDM halos to core-collapse and become denser than DM halos in presence of baryons \citep{2002ApJ...568..475B,2002ApJ...581..777C,2011MNRAS.415.1125K,2012MNRAS.423.3740V,2000ApJ...543..514K}. As long as the baryonic component dominates the central region, core-collapse can occur for $\sigma/m =0.5 $ cm$^2$ g$^{-1}$ \citep{2018ApJ...853..109E}. This is the reason why SIDM model predicts both cored and cuspy profiles, depending on baryon concentration. As a result, the coupling between the SIDM and baryons also provides an explanation for the uniformity of the rotation curves \citep{2019PhRvX...9c1020R,2017PhRvL.119k1102K,2017MNRAS.468.2283C}.

\medskip

Figure~\ref{c3f5} shows that the logarithmic slope of the DM density profile, at 1.5$\%$ of the virial radius inferred from the SIDM fits, is correlated with the stellar mass \citep{2019PhRvX...9c1020R}. Then, SIDM+baryons model with an interaction cross-section of $3 $ cm$^2$ g$^{-1}$ can reproduce galaxy rotation curves from $\sim$ 50 to 300 km s$^{-1}$ \citep{2019PhRvX...9c1020R,2017PhRvL.119k1102K,2019arXiv191100544K}. The slope $\alpha$ of SIDM fits, which include the baryonic impact, spans a large range from -0.5 to -2.5, indicating that the SIDM model predicts both cored and cuspy halos. It was also pointed out that this reflects different baryon distributions in galaxies, which have a large impact on SIDM halos. Thus, the SIDM model predicts cored DM density profiles in low surface brightness galaxies and cuspy density profiles in high surface brightness galaxies. It therefore agrees best with observations. This coupling works because within the characteristic scale of these galaxies, the DM and the baryonic masses are comparable. As halos, that host concentrated stellar populations, exhibit few differences in density profiles between CDM and SIDM models in the presence of baryons, the resulting DM core is effectively indistinguishable between CDM and SIDM (see Figure~\ref{c3f5}). Maybe signatures in stellar kinematics could distinguish between these two core formation mechanisms, one impulsive (feedback) and the other adiabatic (SIDM) \citep{2019MNRAS.485.1008B}. However, the impact of baryonic physics in ultra-faint galaxies is negligible, such that it is difficult to imagine how a population of dense ultra-faint galaxies can be accommodated with a constant cross-section of $\sigma/m = 3 $ cm$^2$ g$^{-1}$ (see Figure~\ref{c3f5}) \citep{2019PhRvD.100f3007Z}. 

\subsubsection{Primordial black holes as dark matter candidates}

\begin{figure}[!t]
\centering
\includegraphics[width=0.75\textwidth]{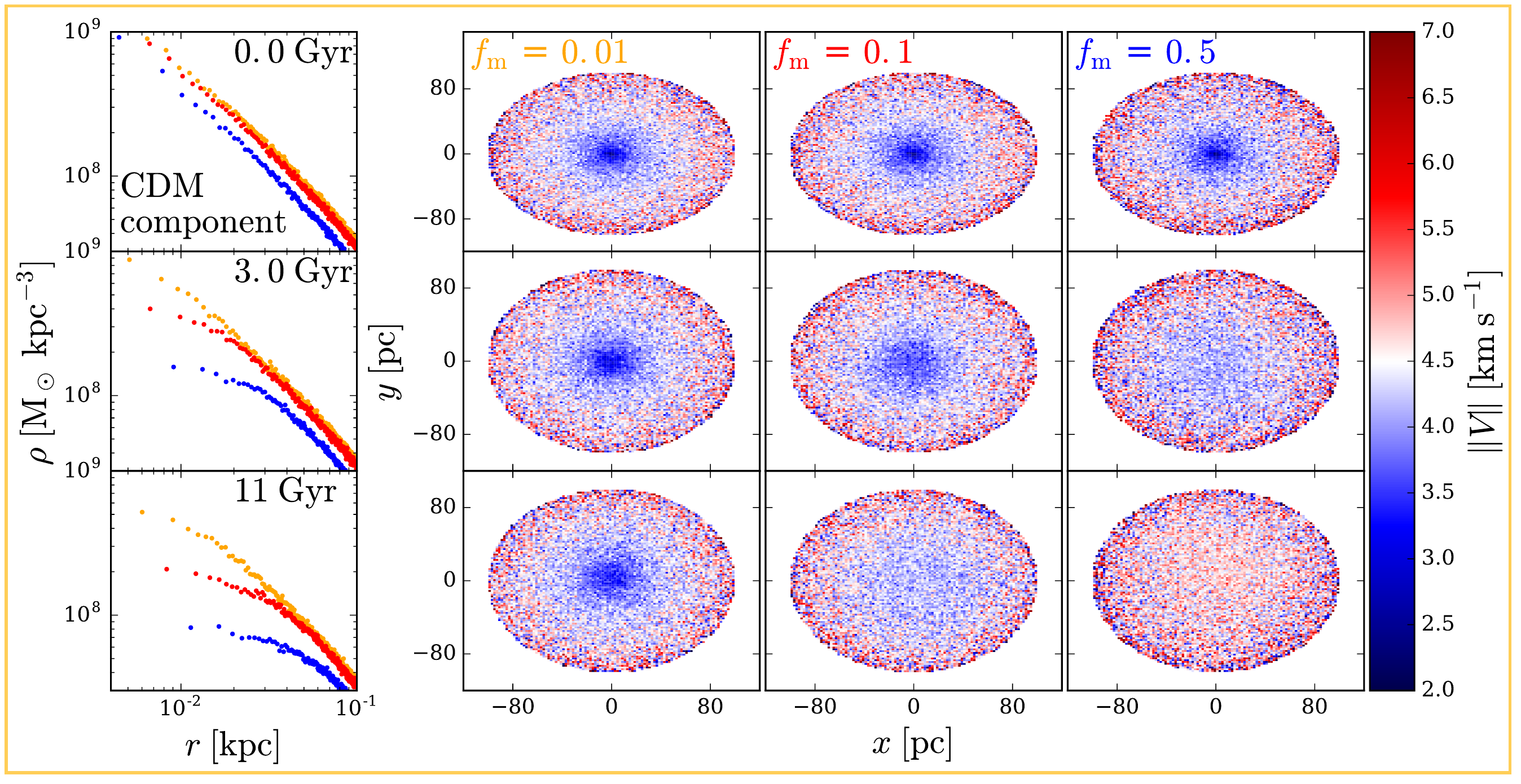}
\caption{{\it Cusp-to-core transition due to PBH-DM heating:} Density profiles of the CDM component ({\it left panels}) and maps of the CDM mass-weighted velocity distribution projected face-on through a 100 pc ({\it right panels}) over the time. The PBH+CDM halos are populated with 100 M$_{\odot}$ PBHs assuming $r_{\mathrm{s}}^{\mathrm{PBH}} = r_{\mathrm{s}}^{\mathrm{CDM}}$ and different $f_\mathrm{m}=0.5$. Core formation goes along with dynamical heating of CDM particles. This figure is reprinted from \cite{2019arXiv191100544K}.}
\label{pbh1}
\end{figure}

Even if weakly interacting massive elementary particles are the most popular DM candidates, DM could be made of macroscopic compact halo objects such as primordial black holes (PBHs) (\cite{1967SvA....10..602Z,1971MNRAS.152...75H,2010RAA....10..495K,2018PDU....22..137C}).These black holes could naturally be produced in the early Universe via cosmic inflation, without the need to appeal to new physics beyond the standard model (\cite{2017PhRvD..96d3504I,2015PhRvD..92b3524C} and for a recent review \cite{2021arXiv211002821C}) . One of the three allowed mass windows around 25 - 100 M$_{\odot}$ is of special interest in view of the recent detection of black-hole mergers by LIGO  \citep{2017PhRvD..96b3514C,2016PhRvL.116f1102A} and could potentially also detected by the Laser Interferometer Space Antenna (LISA) \citep{2017arXiv170200786A}. 

\medskip

The cusp-core problem in $10^7$ M$_{\odot}$ halos such as low-mass dwarf galaxies by considering the possibility that a fraction of the DM is made of PBHs was addressed by \cite{2020MNRAS.492.5218B}. For DM halo composed of CDM particles and PBHs (DM = PBH + CDM), they have defined the PBH+CDM mass fraction as 
\begin{equation}
f_\mathrm{m} = \frac{M_{\mathrm{PBH}}}{M_{\mathrm{CDM}}},
\label{hop}
\end{equation}
where $M_{\mathrm{PBH}}$ and $M_{\mathrm{CDM}}$ are the total masses of PBHs and CDM particles.
It is known that in collisionless systems such as globular clusters, massive stars fall towards the centre of the potential well and their energy is transferred to the lighter stars, which move away from the centre \citep{1969ApJ...158L.139S,1943RvMP...15....1C}. Consequently, the density profile of lighter stars change due to this diffusion process \citep{2016ApJ...824L..31B,2017PhRvL.119d1102K,2018MNRAS.476....2Z} . In the same manner, \cite{2020MNRAS.492.5218B} have demonstrated using high performance $N$-body simulations on GPU that PBHs, as DM candidates, can induce a cusp-to-core transition in PBH+CDM halos through gravitational heating from two principal mechanisms, dynamical friction by CDM particles on PBHs and two-body relaxation between PBH and CDM (see Figure~\ref{pbh1}). As the CDM particle velocity increases in the central region, the CDM density profile changes until core formation occurs. This figure demonstrates that core formation goes along with dynamical heating of CDM particles.

\medskip

They suggest that this core formation mechanism works with a lower limit on the PBH mass fraction of $1\%$ of the total dwarf galaxy dark matter content \citep{2020MNRAS.492.5218B}. This cusp-to-core transition takes between 1 and 8 Gyr to appear, depending on the fraction $f_m$, the PBH mass $m_{\mathrm{PBH}}$ and the PBH scale radius $r_{\mathrm{s}}^{\mathrm{PBH}}$ \citep{2020MNRAS.492.5218B}. As cores occur naturally in PBH+CDM halos without the presence of baryons, there is no cusp-core problem in this alternative theory. However, this mechanism seems only efficient in low mass galaxies as the core formation time is proportional to the halo mass. That is the reason why cores in higher mass galaxies could form only via a hierarchical scenario, in other words through halo mergers. Even if this alternative theory was already investigated in a cosmological context with only DM+PBHs \citep{2019PhRvD.100h3528I}, this mechanism needs to be tested in the presence of baryons.


\newpage
\section{Conclusions}

As understanding how the dark matter (DM) is distributed in the central region of galaxy is directly related to one of the major unsolved problems in astrophysics, the nature of DM, it is not surprising that the cusp-core problem in dwarf galaxies in accordance with observations still remains a challenge. This review was intended to discuss all the main research avenues for solutions to this small-scale issue within cold dark matter (CDM) but also in alternative theories. 

In the future, the cusp-core problem must be approached from two main angles. First, an accurate inference of the DM density profile from observations is necessary. This should become possible with more radial velocities in the central regions of dwarfs thanks to future Gaia data release \citep{2021A&A...649A...1G}. Currently, DM densities are barely constrained observationally at $\sim$100 pc scales. As stressed by \cite{2021MNRAS.507.4715C}, we need more member stars for dwarfs to properly use the Jeans analyse by assuming realistic non-spherical geometry for halos. Then we could focus more in depth on other dwarfs than Fornax, which has been extensively investigated because of its large stellar mass (see Table~\ref{tab1}). 

Second, all core formation mechanisms within CDM need to be addressed in a cosmological context to check their efficiency during the Universe formation but it is also crucial to find observational signatures of these mechanisms in order to distinguish and maybe exclude some of them. For instance, if tidal effects are responsible for DM core formation in our local dwarfs, such tidal tails should be detectable in future surveys \citep{2020arXiv201109482G}. Concerning alternative DM theories, efforts must be pursued in constraining their additional degrees of freedom compared to CDM. Very recently, an new upper limit on the self-interacting scattering cross-section in SIDM universe was imposed based by comparing the measurements of the central density at 150 pc of subhalos in a high-resolution cosmological simulation and our local dwarf galaxies \citep{2021arXiv210705967E}. 

As outlined in the review, the contribution of baryons in a gravitational and hydrodynamical fashion on the distribution of DM within galaxies is non-negligible. However, the presence of baryons can potentially biased our understanding of the DM properties. For instance, both stellar feedback and SIDM can initiate cusp-to-core transformation in dwarf halos. However, it was pointed out these core mechanisms act on different timescales on which they affect the gravitational potential \citep{2019PhRvD.100f3007Z}. Furthermore, it was demonstrated that they could have a distinct signature in the velocity dispersion profiles of stars \citep{2021arXiv210807358B}. Such observable properties can be used to distinguish these two mechanism. 

Future missions such as the James Webb Space Telescope have the ambition to give us a direct insight into DM halos of very high redshift galaxies. These very old DM structures will not have been altered by effects of the environment yet. The presence of DM cores in these galaxies will dramatically favor alternative theory where cores emerge naturally. On the contrary, the absence of cores will reinforce the CDM model and dynamical perturbers such as stellar feedback of infalling structures will be responsible for the formation of cores at low redshift. 

\section*{Acknowledgements}

I thank the three reviewers for their constructive feedback which helped to improve the quality of the manuscript. I thank Joseph Silk for useful comments and suggestions. I also thank Eduardo Vitral for illuminating discussions about Gaia data. 


\reftitle{References}

\end{paracol}
\end{document}